\documentclass[useAMS,usenatbib]{mn2e}
\usepackage{times} 
\usepackage{amsmath}
\usepackage{rotating}
\usepackage{amssymb}
\newcommand{\sm}{\sc}

\title[The fine-grained phase-space structure of Cold Dark Matter halos]
{The fine-grained phase-space structure of Cold Dark Matter halos}
\author[Vogelsberger et al.] 
{
Mark Vogelsberger$^{1}$,
Simon D.M. White$^{1}$,
Amina Helmi$^{2}$, 
Volker Springel$^{1}$\\
$^1$ Max-Planck Institut fuer Astrophysik,
Karl-Schwarzschild Strasse 1, 
D-85748 Garching, 
Germany \thanks{\texttt{\{vogelsma,volker,swhite\}@mpa-garching.mpg.de}} \\
$^2$ Kapteyn Astronomical Institute, 
University of Groningen, 
P.O. Box 800, 
9700 AV Groningen, 
Netherlands \thanks{\texttt{ahelmi@astro.rug.nl}} \\
}

\date{}
\begin{document}

\date{Accepted ???. Received ???; in original form ???}

\pagerange{\pageref{firstpage}--\pageref{lastpage}} \pubyear{2007}

\maketitle

\label{firstpage}

\begin{abstract}
We present a new and completely general technique for calculating the
fine-grained phase-space structure of dark matter throughout the
Galactic halo.  Our goal is to understand this structure on the scales
relevant for direct and indirect detection experiments.  Our method is
based on evaluating the geodesic deviation equation along the
trajectories of individual DM particles. It requires no assumptions
about the symmetry or stationarity of the halo formation process. In
this paper we study general static potentials which exhibit more
complex behaviour than the separable potentials studied previously.
For ellipsoidal logarithmic potentials with a core, phase mixing is
sensitive to the resonance structure, as indicated by the number of
independent orbital frequencies. Regions of chaotic mixing can be
identified by the very rapid decrease in the real space density of the
associated dark matter streams. We also study the evolution of stream
density in ellipsoidal NFW halos with radially varying isopotential
shape, showing that if such a model is applied to the Galactic halo,
at least $10^5$ streams are expected near the Sun.  The most novel
aspect of our approach is that general non-static systems can be
studied through implementation in a cosmological N-body code. Such an
implementation allows a robust and accurate evaluation of the
enhancements in annihilation radiation due to fine-scale structure
such as caustics.  We embed the scheme in the current state-of-the-art
code GADGET-3 and present tests which demonstrate that N-body
discreteness effects can be kept under control in realistic
configurations.
\end{abstract}



\section{Introduction} \label{sect:intro}

Dark Matter (DM) is still to be directly detected. The first indirect
indications of its existence came in the 1930s, with measurements of
the velocities of galaxies in clusters. The cluster mass required to
gravitationally bind the galaxies was found to be more than an order
of magnitude larger than the sum of the luminous masses of the
individual galaxies \citep{1933AcHPh...6..110Z,1936ApJ....83...23S}.
The early detection of large amounts of unseen matter associated with
the Local Group \citep{1959ApJ...130..705K} was followed in the
1970s by observations of the rotation curves of spiral galaxies which
showed that these are flat, or even rising, at distances far beyond
their stellar components
\citep{1970ApJ...159..379R,1979ARA&A..17..135F,1980ApJ...238..471R}.
Studies of satellite systems suggested that the mass distributions of
most galaxies might be an order of magnitude larger and more massive
than their visible parts
\citep{1974ApJ...193L...1O,1974Natur.250..309E}.  All these
discoveries led to the conclusion that a large fraction of mass in the
Universe is dark. This has also been supported by recent gravitational
lensing studies that demonstrate the existence of extended massive DM
halos (e.g. Mandelbaum et al. 2006).

As the dominant mass component of galaxies and large-scale structures,
DM has necessarily become a key ingredient in theories of cosmic
structure formation. The most successful of these theories is the
hierarchical paradigm.  In the current version of this model, the DM
is composed of nonbaryonic elementary particles known as Cold Dark
Matter (CDM) \citep{1982ApJ...263L...1P}.  The term ``cold" derives
from the fact that the DM particles had negligible thermal motions at
the time of matter-radiation equality. Their abundance was set when
the interaction rate became too small for them to remain in thermal
equilibrium with other species in the expanding Universe. The first
objects to form in a CDM Universe are small galaxies. They then merge
and accrete to give rise to the larger structures we observe
today. Thus structure formation occurs in a ``bottom-up" fashion
\citep{1984Natur.311..517B,1985ApJ...292..371D,2006Natur.440.1137S}.

The crucial test of this paradigm undoubtedly consists in the
determination of the nature of DM through direct detection
experiments. Among the most promising candidates from the particle
physics perspective are axions and neutralinos.  Axions were
introduced to explain the absence in Nature of strong-CP (Charge
conjugation and Parity) violations \citep{1977PhRvL..38.1440P}. The
cosmological population formed out of equilibrium as a zero momentum
Bose condensate.  They can be detected through their conversion to
photons in the presence of a strong magnetic field, 
(e.g. Ogawa et al. 1996; Hagmann et al. 1998). Neutralinos are the
lightest supersymmetric particles, and may be considered as a
particular form of weakly interacting massive particles (WIMPs)
\citep{1985NuPhB.253..375S}. The most important direct detection
process for neutralinos is through elastic scattering on nuclei.

Today many experiments are searching for these particles,
\citep{2004PhRvL..93u1301A, 2005PhRvD..71l2002S, 2006astro.ph.12565S,
2007JPhCS..60...58A, 2007arXiv0705.3345S}.  The main challenge lies in
the large background they encounter.  Having an idea what the detector
signal might look like can help substantially in fine-tuning the
experiments in order to increase the chance of a detection. In
addition, many experiments are attempting to detect WIMPs indirectly
by searching for $\gamma$-ray emission from their annihilation 
\citep{2005NewAR..49..213D,2005PhRvL..95t9001D,2006PhRvD..73f3510B,2007JCAP...06...13H}.
 Predictions for this radiation are currently uncertain
because very substantial enhancements are possible, at least in
principle, from fine-scale structure such as caustics in the dark
matter distribution \citep{2001ApJ...561...35D,2007JCAP...05...15M}.

The differential detection rate for WIMPs in a given material (per
unit detector mass) is sensitive to the local velocity distribution:
\begin{equation}
\frac{\mathrm{d}R}{\mathrm{d}Q} = \frac{\rho}{M_\chi} \int d^3 v f(\mathbf{v}) v \frac{\mathrm{d}\sigma}{\mathrm{d}Q}
\label{eq:DIFF_WIMP_DET_RATE}
\end{equation}
where $Q$ is the energy deposited in the detector and
$\mathrm{d}\sigma/\mathrm{d}Q$ is the differential cross section for
WIMP elastic scattering with the target nucleus. $M_\chi$ is the WIMP
mass which lies in the GeV to TeV range for currently preferred
models. $\rho$ is the local mass density and $f(\mathbf{v})$ the
velocity distribution of WIMPs that reach the detector.

From Eq. (\ref{eq:DIFF_WIMP_DET_RATE}) it is evident that the count
rate in a direction-insensitive experiment depends on the velocity
distribution of the incident particles and will be modulated by the
orbital motion of the Earth around the Sun
\citep{1986PhRvD..33.3495D}. In most studies, an isotropic Maxwellian
distribution relative to the galactic halo has been assumed,
(e.g. Freese et al. 1988), although there are other models in
the literature, discussing, for example, multivariate Gaussians
\citep{2000MNRAS.318.1131E}.  Some attempts at understanding the
effect of fine-scale structure in the velocity distribution of DM
particles have also been made
\citep{1998PhLB..432..139S,2001PhRvD..64h3516S,2001PhRvD..64f3515H}.

\begin{figure}
\centerline{\includegraphics[width=1.0\linewidth]{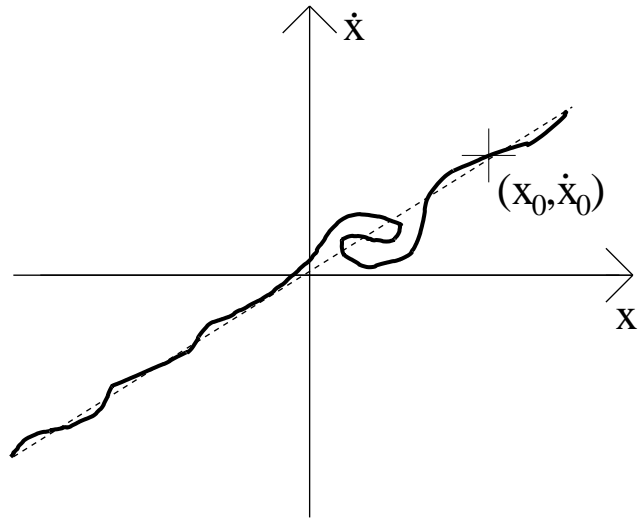}}
\caption{
\label{fig:FLOW}
Sketch of an idealised CDM phase-space sheet in the $x, \dot{x}$
plane. The thickness of the line represents the local velocity
dispersion within each stream. The small wiggles correspond to initial
density perturbations and the multi-valued region reflects the
multiple streams created by winding in non-linear regions.  Depending
on $x$ position an observer sees one or three streams. At points where
the number of streams changes by two, a caustic with a very high DM
density is present. Such caustics may be significant for the total
annihilation flux. The number of streams at each point is a measure
of the local amount of mixing. The dashed line represents the Hubble
Flow. The cross marks the phase-space coordinates of a particular CDM
particle embedded in the flow.  }
\end{figure}

A significant signal could come from what are known as streams of DM
\citep{1995PhRvL..75.2911S,2002PhRvD..66f3502H,2005PhRvD..72h3513N}. Both
axions and WIMPs are cold.  In the absence of clustering their present
day velocity dispersion would be negligible ($\delta v \sim 10^{-10}c$
for WIMPs and $\delta v \sim 10^{-17}c$ for axions).  They are
effectively restricted to a 3D hypersurface, a sheet in 6D
phase-space.  The growth of structure results in continual stretching
and folding in phase-space of this initially almost uniform sheet.
This process is called mixing. The more strongly the system mixes, the
more streams of DM particles are present at a given location in
configuration-space. Mixing stretches each sheet and so its density
decreases.  The existence of distinct streams is a direct consequence
of the collisionless character and the coldness of CDM.  The situation
is sketched in Fig. \ref{fig:FLOW}. At the points where the number of
streams changes by two, the local configuration-space density of dark
matter becomes extremely high. These are caustics of the kind studied
by catastrophe theory \citep{0471050644,1999MNRAS.307..877T}.  Note
that Liouville's Theorem prevents the CDM phase-space sheet from ever
tearing, although it can be arbitrarily strongly stretched.

If the dark matter density in the solar neighbourhood is dominated by
a small number of streams, its velocity distribution will effectively
consist of a few discrete values, one for each stream.  If, on the
other hand, the number of streams is very large, the velocity
distribution will be smooth and individual streams will be
undetectable. This issue has so far been addressed only under
simplified conditions and divorced from its proper cosmological
context \citep{2002PhRvD..66f3502H,2005PhRvD..72h3513N}.  This is
largely because the only tool capable of studying cosmological
structure formation in full generality, namely N-body simulations,
cannot resolve the relevant scales. For example,
\cite{2005PhRvD..72h3513N} estimate that of the order of $10^{12}$
particles would be required in a simulation of the Milky Way's halo to
begin to resolve the streams in the solar neighbourhood.  Even the
largest simulation so far published, the Via Lactea model
\citep{2007ApJ...657..262D}, is 4 orders of magnitude short of this
minimum requirement. It is thus impossible to figure out the
number of streams near the Sun and the properties of ``typical'' 
streams with current N-body capabilities.

A related issue that has been much discussed is whether caustics can
affect direct or indirect detection experiments. Some authors claim
that caustics play an important role as stable phenomena connected to
any collapsing CDM halo and taking the form of relatively massive
rings or shells \citep{1999PhRvD..60f3501S,
2006PhRvD..73b3510N}.  Other authors have argued that caustics should
be more abundant, weaker and dynamically negligible
\citep{2002PhRvD..66f3502H}. Even if caustics negligibly affect the
gravitational potential, they may have very substantial effects on the
annihilation rate of dark matter \citep{2001PhRvD..64f3515H,
2001PhRvD..63h3515B, 2005JCAP...05..007P, 2006MNRAS.366.1217M,
2007JCAP...05...15M, 2007PhRvD..75l3514N}. All these papers were able
to evaluate the enhancements due to caustics only under restrictive
and unrealistic assumptions about symmetry, formation history, etc.
While they demonstrate that large enhancement factors may be possible,
they do not provide reliable estimates for the actual enhancement
expected.  The method we present below is capable of providing such
estimates for halos growing as predicted by the $\Lambda$CDM model.

In this paper we will present a novel approach that directly analyses
structure in the fine-grained phase-space distribution.  We circumvent
the ``number of particle" problem by solving the geodesic deviation
equation (GDE) for every DM particle.  This allows us to calculate the
local properties of the DM stream each simulation particle is embedded
in, in particular, its configuration-space density and its local
velocity distribution. The mass-weighted number of streams near any
point is then the total local density divided by the mean density of
the individual local streams. Caustic passages can be detected
robustly from the properties of each particle's local stream, and the
particle's instantaneous annihilation probability within this stream
can be evaluated simply and integrated accurately through caustics.

This paper is the first of a series. Here we introduce our method and
apply it to study the evolution of streams in static potentials that
are too complex to be analysed using previous techniques.  We also
implement our scheme in a state-of-the-art N-body simulation code, and
use simple test problems to demonstrate that N-body discreteness
effects can be kept under control in realistic applications. Later
papers will address issues associated with mixing, caustics and
annihilation radiation in the full cosmological context.

The outline of our paper is the following: In Section~2 we present a
detailed derivation of the GDE and show how it can be used to quantify
mixing, to locate caustics, and to calculate stream densities and
annihilation rates. Section~3 describes our code, DaMaFlow, which is
designed to solve the GDE for single orbits in a wide variety of
potentials. In Section~4 we analyse static, separable potentials and
compare results from our method to previous work. The fifth section
applies our scheme to non-integrable, but still static potentials,
revealing their complex phase-space structure.  In Section~6 we turn
to more realistic non-spherical CDM halo potentials and discuss the
influence of halo shape on stream density behaviour. Section~7
discusses the implementation of our method in an N-body code and
presents results of simple tests of when discreteness effects
compromise studies of stream densities and caustics. The final
section summarises our results and gives some conclusions.

\section{The geodesic deviation equation} \label{sect:nong}

Our scheme for calculating the evolution of the fine-grained
phase-space distribution in the neighbourhood of a DM particle is
based on the evolution of the distance between two infinitesimally
separated particle trajectories. This evolution is described by the
Geodesic Deviation Equation (GDE). We use the following notation to
clearly distinguish between three- and six-dimensional quantities: an
underline denotes a $\mathbb{R}^3$ vector and two of them denote a
$\mathbb{R}^{3 \times 3}$ matrix. An overline denotes a $\mathbb{R}^6$ vector and two of
them denote a $\mathbb{R}^{6 \times 6}$ matrix. Thus a general
phase-space vector is composed of two three-dimensional vectors:
$\overline{x}=(\underline{x},\underline{v})$.

To derive the GDE we first write down the equations of motion for 
a DM particle. These are simply
\begin{equation}
\ddot{\underline{x}}\left (t;\underline{x}_0, \underline{v}_0\right ) = - \underline{\nabla}_{x} \Phi\left (\underline{x}\left (t;\underline{x}_0, \underline{v}_0\right )\right ),
\label{eq:NEWTONIAN_MOTION}
\end{equation}
with initial conditions $\underline{x}\left (t_0;\underline{x}_0, \underline{v}_0 \right ) = \underline{x}_0 \quad
\underline{v}\left (t_0;\underline{x}_0, \underline{v}_0 \right ) = \underline{v}_0$.

As $\dot{\underline{x}}=\underline{v}$  the equation of motion in
phase-space can be written:
\begin{equation}
\dot{\overline{x}}\left(t;\overline{x}_0\right) =
\left(
   \begin{array}{c}
     \underline{v}\\
     - \underline{\nabla}_{x} \Phi\left(\underline{x}\left(t;\overline{x}_0\right) \right) \\
   \end{array}
\right)  = \overline{\Psi}\left(\overline{x}\left(t;\overline{x}_0\right) \right),
\label{eq:PHAE_SPACE_MOTION}
\end{equation}
with initial conditions $\overline{x}\left(t_0;\overline{x}_0\right)=\overline{x}_0=(\underline{x}_0,\underline{v}_0)$.

We want to calculate the local stream density around the DM particle whose trajectory in phase-space is given by
$\overline{x}\left (t;\overline{x}_0\right )$. To do so, we first ask how the displacement vector to a neighbouring DM particle
in phase-space evolves with time:
\begin{equation}
 \overline{\delta}\left (t\right ) = \overline{x}\left (t; \overline{x}_0+\overline{\delta}_0\right ) - \overline{x}\left (t;\overline{x}_0\right ).
 \label{eq:DELTA_DISPLACEMENT}
\end{equation}
Note that $\overline{\delta}\left (t\right )$ is the displacement
between the reference DM particle, which was at $\overline{x}_0$ at
time $t_0$, and another particle which was at $\overline{x}_0 +
\overline{\delta}_0$ at $t_0$.  We are interested in properties in the
immediate neighbourhood of the reference particle, so
$\overline{\delta}_0$ is an infinitesimal displacement.  This allows
us to work to linear order:
\begin{equation}
 \overline{\delta}\left (t\right )  \cong \left (\overline{\delta}_0 \cdot \overline{\nabla}_{x_0}\right ) \overline{x}\left (t;\overline{x}_0\right ).  
 \label{eq:LINEAR_DIFF}
\end{equation}
Introducing the phase-space distortion tensor $\overline{\overline{D}}$ (note that this is a $6\times6$ tensor)
\begin{equation}
 \overline{\overline{D}}\left (t; \overline{x}_0\right ) \equiv \frac{\partial \overline{x}}{\partial \overline{x}_0} \left (t; \overline{x}_0\right ),
 \label{eq:DISTORTION_TENSOR}
\end{equation}
we can rewrite Eq. (\ref{eq:LINEAR_DIFF}) as a simple linear transformation from the starting phase-space 
displacement $\overline{\delta}_0$ to the displacement $\overline{\delta}(t)$ at time $t$:
\begin{equation}
 \overline{\delta}\left (t\right ) \cong \overline{\overline{D}}\left (t;\overline{x}_0\right ) \overline{\delta}_0.
 \label{eq:DELTA_TIME_EVOLUTION}
\end{equation}
Because $\overline{\delta}_0$ is an arbitrary displacement vector, the
distortion tensor describes how the complete local phase-space
neighbourhood around the reference DM particle gets distorted while it
is orbiting in the given potential.  The time evolution of
$\overline{\delta}(t)$ follows from the time evolution of the two
trajectories.  Again we can work this out in linear order\footnote{As
$\overline{\delta}_0$ can be chosen arbitrarily small, this is always
possible.}:
\begin{eqnarray}
 \dot{\overline{\overline{D}}}\left (t;\overline{x}_0\right ) \overline{\delta}_0 \nonumber &\cong&  
 \dot{\overline{\delta}} \left (t\right ) \\ \nonumber
 &=&
 \overline{\Psi}\left(\overline{x}\left(t;\overline{x}_0\right) + \overline{\delta}\left(t\right)\right)
 -\overline{\Psi}\left(\overline{x}\left(t;\overline{x}_0\right)\right) \\ \nonumber
 &\cong&
 \left (\overline{\delta}\left (t\right ) \cdot \overline{\nabla}_{x}\right ) \overline{\Psi}\left (\overline{x}\left (t;\overline{x}_0\right )\right )\\ 
 &\cong&
\left (\left (\overline{\overline{D}}\left (t;\overline{x}_0\right ) \overline{\delta}_0\right ) \cdot \overline{\nabla}_{x}\right ) \overline{\Psi} \left (\overline{x}\left (t;\overline{x}_0\right )\right ). 
 \label{eq:DISTORTION_EQ_MOTION_TAYLOR}
\end{eqnarray}
To derive the equation of motion for the distortion tensor itself we evaluate Eq. (\ref{eq:DISTORTION_EQ_MOTION_TAYLOR}) for
six unit vector phase-space displacements $\overline{\delta}^{(j)}_0$ with $\delta_{0,\alpha}^{(j)}=\delta_{\alpha j}$ 
where $\alpha,j=1,2,\ldots,6$,  and $\delta_{ab}$ is the Kronecker delta. Taking into account Einstein's sum convention this 
yields for each component of Eq. (\ref{eq:DISTORTION_EQ_MOTION_TAYLOR})
\begin{eqnarray}
  \dot{D}_{i j}\left (t;\overline{x}_0\right )   \nonumber
  &=&
  \dot{D}_{i\alpha}\left (t;\overline{x}_0\right ) \delta_{0,\alpha}^{(j)} \\ \nonumber
  &\cong&  \left (D_{\beta \gamma}\left (t;\overline{x}_0\right ) \delta_{0,\gamma}^{(j)} \frac{\partial}{\partial x_\beta}\right ) \Psi_i\left (\overline{x}\left (t; \overline{x}_0\right )\right ) \\ \nonumber
  &=&  \left (D_{\beta \gamma}\left (t;\overline{x}_0\right ) \delta_{\gamma j} \frac{\partial}{\partial x_\beta}\right ) \Psi_i\left (\overline{x}\left (t; \overline{x}_0\right )\right ) \\ \nonumber  
  &=&  \left (D_{\beta j}\left (t;\overline{x}_0\right ) \frac{\partial}{\partial x_\beta}\right ) \Psi_i\left (\overline{x}\left (t; \overline{x}_0\right )\right ) \\ 
  &=&  T_{i \beta} \left(t;\overline{x}_0\right) D_{\beta j}\left (t;\overline{x}_0\right ), 
 \label{eq:DISTORTION_COMPONENT}
\end{eqnarray}
where we have introduced the phase-space tidal tensor
\begin{equation}
\overline{\overline{T}}\left(t;\overline{x}_0\right) = \left(
   \begin{array}{cc}
     \underline{\underline{0}} & \underline{\underline{1}} \\
     \underline{\underline{T}}\left(t;\underline{x}_0\right) & \underline{\underline{0}} \\
   \end{array}
\right),
\label{eq:TIDAL_TENSOR}
\end{equation}
and $\underline{\underline{T}}$ is the configuration-space tidal
tensor given by the second derivatives of the gravitational potential
$T_{i j}=-\partial^2 \Phi/ \partial{x_i} \partial{x_j} $.  As we are
are only interested in linear order we replace $\cong$ by $=$ in
Eq. (\ref{eq:DISTORTION_COMPONENT}) and get an equation of motion for
the 6D distortion tensor:
\begin{equation}
 \dot{\overline{\overline{D}}}\left (t; \overline{x}_0\right ) = \overline{\overline{T}}\left (t; \overline{x}_0\right ) \overline{\overline{D}} \left (t; \overline{x}_0\right ).
 \label{eq:DISTORTION_EQ_MOTION}
\end{equation}
Note that this first-order tensor differential equation represents a
system of 36 coupled ordinary first-order differential equations. To
solve them we need to specify initial conditions. These follow from
the constraint
\begin{equation}
 \overline{\delta}_0 = \overline{\overline{D}}\left (t_0; \overline{x}_0\right ) \overline{\delta}_0 \quad \Rightarrow \quad \overline{\overline{D}}\left (t_0; \overline{x}_0\right ) = \overline{\overline{1}}.
 \label{eq:DISTORTION_IC}
\end{equation}
DM behaves like a collisionless fluid and its fine-grained phase-space
density $f(\overline{x},t) $ is described by the well-known Vlasov
equation:
\begin{equation}
\frac{\partial f}{\partial t} + \underline{v} \cdot \underline{\nabla}_{x} f - \underline{\nabla}_{x} \Phi \cdot \underline{\nabla}_{v} f = 0,
 \label{eq:VLASOV}
\end{equation}
Using the Lagrangian derivative this reads
$\mathrm{D}f/\mathrm{D}t=0$, meaning that the local fine-grained
phase-space density has to be conserved along the orbit of every
particle in the system.  Imagine $N$ particles that fill a small
phase-space volume
$\mathrm{d}V_0=\mathrm{d}\underline{x}_0\mathrm{d}\underline{v}_0$
around the DM reference particle at time $t_0$. At a later time $t$
these particles fill a volume
$\mathrm{d}V=\mathrm{d}\underline{x}\mathrm{d}\underline{v}$ around
the reference particle at $\overline{x}(t;\overline{x}_0)$.
Conservation of phase-space density implies that the two volumes are the same
$\mathrm{d}V_0=\mathrm{d}V$. The evolution of the initial displacement
vectors $\overline{\delta}_0^{(n)}$ from the reference particle to one
of the $N$ other particles in the volume is described by the
distortion tensor associated with the reference particle:
\begin{equation}
 \overline{\delta}^{(n)}\left (t\right ) = \overline{\overline{D}}\left (t;\overline{x}_0\right ) \overline{\delta}_0^{(n)} \qquad n=1,2,\ldots,N
 \label{eq:CLOUD}
\end{equation}
The change in volume due to this linear transformation is given by the
determinant $\mathrm{det}(\overline{\overline{D}}\left
(t;\overline{x}_0\right ))$. As this volume has to be conserved, the
determinant of the phase-space distortion tensor has to be
conserved. Note that not only must the volume be conserved, but also
the orientation of the volume element. This means that the sign of the
determinant is also fixed.  From the initial conditions
Eq. (\ref{eq:DISTORTION_IC}) one gets
$\mathrm{det}(\overline{\overline{D}}\left (t;\overline{x}_0\right
))$=1 at all times. This fact can be used to check numerical
solutions of the equations.

The structure of Eq. (\ref{eq:DISTORTION_EQ_MOTION}) allows the
equations of motion for the distortion to be broken down to a set of
equations that is more convenient to work with. Let us first rewrite
Eq. (\ref{eq:DISTORTION_EQ_MOTION}) using blocks of $3\times 3$
tensors:\footnote{We suppress the argument $t;\overline{x}_0$ to avoid
confusion.}
\begin{eqnarray}
\frac{\mathrm{d}}{\mathrm{d}t}
 \left(
   \begin{array}{cc}
     \underline{\underline{D}}_{xx} & \underline{\underline{D}}_{xv} \\ \nonumber
     \underline{\underline{D}}_{vx} & \underline{\underline{D}}_{vv} \\ 
   \end{array}
\right)
&=& \left(
   \begin{array}{cc}
     \underline{\underline{0}} & \underline{\underline{1}} \\ 
     \underline{\underline{T}} & \underline{\underline{0}} \\ 
   \end{array}
\right)
 \left(
   \begin{array}{cc}
     \underline{\underline{D}}_{xv} & \underline{\underline{D}}_{xv} \\ 
     \underline{\underline{D}}_{vx} & \underline{\underline{D}}_{vv} \\ 
   \end{array}
\right) \\ \nonumber
&=&
 \left(
   \begin{array}{cc}
     \underline{\underline{D}}_{vx} & \underline{\underline{D}}_{vv} \\
     \underline{\underline{T}} \quad \underline{\underline{D}}_{xx} & \underline{\underline{T}} \quad \underline{\underline{D}}_{xv} \\
   \end{array}
\right). \nonumber
\label{eq:6D_block}
\end{eqnarray}
Writing down the equation for each matrix block yields four equations:
\begin{equation}
\dot{\underline{\underline{D}}}_{xx} = \underline{\underline{D}}_{vx} \qquad ; \qquad \dot{\underline{\underline{D}}}_{xv} = \underline{\underline{D}}_{vv},
\label{eq:6D_blocks1}
\end{equation}
and
\begin{equation}
\dot{\underline{\underline{D}}}_{vx} = \underline{\underline{T}} \quad \underline{\underline{D}}_{xx} \qquad ;\qquad \dot{\underline{\underline{D}}}_{vv} = \underline{\underline{T}} \quad \underline{\underline{D}}_{xv} .
\label{eq:6D_blocks2}
\end{equation}
These can be combined to give:
\begin{equation}
\ddot{\underline{\underline{D}}}_{xx} = \underline{\underline{T}} \quad \underline{\underline{D}}_{xx} \qquad ; \qquad \ddot{\underline{\underline{D}}}_{xv} = \underline{\underline{T}} \quad \underline{\underline{D}}_{xv} .
\label{eq:6D_second}
\end{equation}
Thus we get two identical differential equations of second-order for
two $3\times 3$ tensors whose dynamics is driven by the ordinary tidal
tensor. From the initial condition for the 6D distortion one can see
that the only difference between $\underline{\underline{D}}_{xx}$ and
$\underline{\underline{D}}_{xv}$ lies in the appropriate initial
conditions:
$\underline{\underline{D}}_{xx}(t_0;\underline{x}_0)=\underline{\underline{1}}$,
$\,\dot{\underline{\underline{D}}}_{xx}(t_0;\underline{x}_0)=\underline{\underline{0}}$
and
$\underline{\underline{D}}_{xv}(t_0;\underline{x}_0)=\underline{\underline{0}}$,
$\,\dot{\underline{\underline{D}}}_{xv}(t_0;\underline{x}_0)=\underline{\underline{1}}$
in the two cases.
From the solutions of these two initial condition problems the 6D
distortion solution can then be constructed:
\begin{equation}
\overline{\overline{D}} = 
 \left(
   \begin{array}{cc}
     \underline{\underline{D}}_{xx} & \underline{\underline{D}}_{xv} \\ 
     \mathrm{d}/\mathrm{d}t\underline{\underline{D}}_{xx} & \mathrm{d}/\mathrm{d}t\underline{\underline{D}}_{xv} \\ 
   \end{array}
\right) .
\label{eq:6D_FULL_EQUATION}
\end{equation}

Up to this point we have worked out all equations in phase-space. As
we are interested in the stream density in configuration-space, we
need to project down to this space. As already mentioned, CDM lies on
a thin sheet in phase-space. This sheet has a certain orientation at
the starting point of the reference DM particle. Take
$(\underline{x},\underline{v}):\underline{v}=\underline{V}(\underline{x};t_0,
\overline{x}_0)$ to be the local parametrisation of the sheet
surrounding this particle at time $t_0$\footnote{Such a
parametrisation is always possible locally, but due to mixing there
is, in general, no simple global parametrisation of the stream. This
is only possible for very early times, where the sheet is dominated by
the Hubble Flow, the $\underline{x}-\underline{v}$ relation is
one-to-one, and the stream density is almost uniform.}.  Now we ask
how an infinitesimal displacement $\underline{\delta}_{0,x}$ in
configuration-space is distorted by evolution. First we note that any
displacement in $\underline{x}$ implies a displacement in
velocity-space due to the restriction of particles to the sheet:
\begin{equation}
\underline{\delta}_{0,v} = \underline{\underline{V}}_x \left(\overline{x}_0\right)  \underline{\delta}_{0,x} \quad ; \quad
\underline{\underline{V}}_x \left(\overline{x}_0\right)= 
\frac{\partial \underline{V}}{\partial \underline{x}}
\left(\underline{x}_0; t_0, \overline{x}_0\right) .
\label{eq:IC_CDM}
\end{equation}
The phase-space distortion $\overline{\overline{D}}$ describes how the
corresponding phase-space displacement
$(\underline{\delta}_{0,x},\underline{\delta}_{0,v})$ evolves. We are
here interested in the configuration-space part of the phase-space
displacement at time $t$:
\begin{equation}
\underline{\delta}_{x}\left(t\right) = \underline{\underline{D}}_{xx}\left(t;\overline{x}_0\right )  \underline{\delta}_{0,x} + 
\underline{\underline{D}}_{xv}\left(t;\overline{x}_0\right )  \underline{\underline{V}}_x\left(\overline{x}_0\right)  \underline{\delta}_{0,x}  .
\label{eq:3D_DELTA}
\end{equation}
The evolution of the displacement in configuration-space can also be
described by a linear transformation:
\begin{equation}
\underline{\delta}_{x}\left(t\right) = \underline{\underline{D}}\left(t;\overline{x}_0\right ) \underline{\delta}_{0,x}, \\  \nonumber
\label{eq:3D_DISTORTION_EQ}
\end{equation}
where we have introduced the configuration-space distortion tensor (note that this is a $3\times 3$ tensor)
\begin{equation}
\underline{\underline{D}}\left(t;\overline{x}_0\right ) = 
\underline{\underline{D}}_{xx}\left(t;\overline{x}_0\right )  + 
\underline{\underline{D}}_{xv}\left(t;\overline{x}_0\right )  \underline{\underline{V}}_x\left(\overline{x}_0\right) .
\label{eq:3D_DISTORTION}
\end{equation}
This tensor can also be derived with the help of two projection operators:
\begin{equation}
\underline{\underline{D}}\left(t;\overline{x}_0\right )=
\left(\underline{\underline{1}} \quad  \underline{\underline{0}} \right) 
\overline{\overline{D}}\left(t;\overline{x}_0\right )
\left(
   \begin{array}{c}
     \underline{\underline{1}} \\
     \underline{\underline{V}}_x\left(\overline{x}_0\right)
   \end{array}
\right) .
\end{equation}
As in the case of phase-space distortion, the change in volume due to
the linear transformation in Eq. (\ref{eq:3D_DISTORTION_EQ}) is given
by the determinant, so that the stream density in configuration-space
is proportional to the inverse of this determinant:
\begin{equation}
\rho_{\mathrm{stream}} \left(t\right) \propto \frac{1}{\left| \mathrm{det}\left(\underline{\underline{D}}\left(t;\overline{x}_0\right )\right)\right|} .
\label{eq:FLOW_DENSITY}
\end{equation}
At time $t_0$ the configuration-space distortion tensor equals
unity. Thus, if we norm the stream density to its initial value, we
get the following relation for the normed stream density:
\begin{equation}
\rho_{\mathrm{stream}}^{\mathrm{normed}} \left(t\right) = \frac{1}{\left| \mathrm{det}\left(\underline{\underline{D}}\left(t;\overline{x}_0\right )\right)\right|} .
\label{eq:NORMED_FLOW_DENSITY}
\end{equation}
In the rest of this paper we will almost always discuss this normed
stream density.

\begin{figure}
\centerline{\includegraphics[width=1\linewidth]{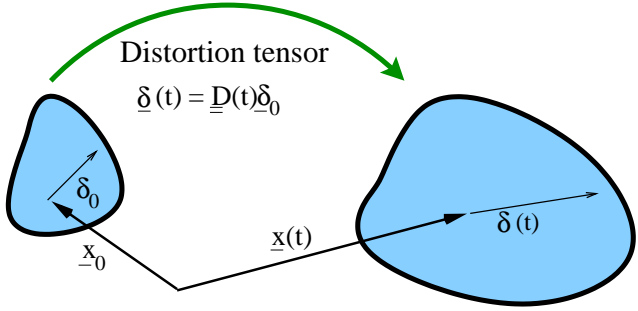}}
\caption{
\label{fig:DISTORTION_FIG}
The configuration-space distortion tensor $\underline{\underline{D}}$
describes how an initial small configuration-space displacement
$\underline{\delta}_0$ evolves in time. This reflects the stretching
of an infinitesimally small cloud of virtual particles around the
reference particle that is placed at $\underline{x}_0$ at time $t_0$.
The stretching of the cloud is driven by the tidal field that the DM
particle encounters as it orbits.  }
\end{figure}

In Fig. \ref{fig:DISTORTION_FIG} we sketch the distortion of the
infinitesimal cloud around the reference DM particle. Note the
difference between the stream density evolution in configuration-space
and in phase-space. The volume of the small cloud grows in
Fig. \ref{fig:DISTORTION_FIG} and is not constant anymore! This is a
result of the projection from phase-space to configuration-space. 
Nearby trajectories spatially diverge in time. 

This formalism can be used to identify caustics in a very efficient
way. If the DM particle passes through a caustic
$\mathrm{det}(\underline{\underline{D}})$ passes through zero, and the
stream density goes to infinity (for perfectly cold DM).  A small
cloud surrounding the reference particle turns inside out as it passes
through the caustic. This corresponds to a change in sign of the
determinant, and thus can easily identified numerically. This property
allows the location of caustics to be mapped accurately even in
complex configurations. Note that the possibility of sign changes
explains why we took the modulus of the determinant in
Eq. (\ref{eq:FLOW_DENSITY}).

The complex fine-grained phase-space structure and especially the
caustics expected in CDM halos are likely to substantially enhance
annihilation radiation. These effects have so far been analysed only
under quite simplified conditions
\citep{2001PhRvD..64f3515H,2001PhRvD..63h3515B,
2005JCAP...05..007P,2006MNRAS.366.1217M, 2007JCAP...05...15M,
2007PhRvD..75l3514N}.  As a result, it is unclear how strong such
enhancement effects will be in the proper cosmological
context. Previous studies of annihilation radiation from N-body halos
had realistic formation histories but were unable to resolve
caustics. They estimated emissivities from the local mean CDM density,
thus effectively excluding contributions from single streams
\citep{2003MNRAS.345.1313S,2007ApJ...657..262D}. \cite{2001PhRvD..64f3515H}
noted that this results in an underestimation of the annihilation
rate, and suggested that annihilation might, in fact, be dominated by
contributions from the neglected caustics (which he baptised as
``micropancakes").  

Our formalism enables a robust and accurate calculation of the
contribution to the annihilation radiation from individual
streams. The annihilation rate for each particle due to encounters
with other particles in its own stream can be evaluated directly from
the local stream velocity distribution and density. Integrating these
rates along the trajectories of all particles produces a Monte Carlo
estimate of the intrastream annihilation rate for the system as a
whole. This automatically includes the contributions from all caustics
and is exactly the fine-scale contribution which is missing from
standard N-body-based estimates of annihilation luminosities.

We now discuss briefly how this is done.  Given the very small
primordial velocity dispersion $\sigma_0$ of CDM, and approximating
the initial configuration-space density as a constant $\rho_0$ we can
write the phase-space density around a reference particle at the
initial time $t_0$ as follows:
\footnote{$\dagger$ denotes the transpose of a matrix}
\begin{equation}
 f\left(\overline{x},t_0\right)  =  \rho_0 N_0 \exp(-\frac{1}{2}(\overline{x}-\overline{x}\left(t_0;\overline{x}_0\right))^{\dagger} \overline{\overline{W}}_0 (\overline{x}-\overline{x}\left(t_0;\overline{x}_0\right))), \nonumber
 \label{eq:GAUSSIAN} \nonumber
\end{equation}
where
\begin{equation}
\overline{\overline{W}}_0=\sigma_0^{-2}\mathrm{diag}(\underline{\underline{0}},\,\underline{\underline{1}}), \nonumber
 \label{eq:DISPERIONS}
\end{equation}
and $N_0=(2\pi)^{-3/2} \sigma_0^{-3}$.  Note that this represents a
Gaussian distribution in velocity-space and a constant density in
configuration-space.

Using $\overline{\delta}(t) = \overline{\overline{D}}(t)
\overline{\delta}_0$ we obtain the phase-space density around the
particle at the later time $t$:
\begin{equation}
 f\left (\overline{x},t\right)  = \rho_0 N_0 \exp(-\frac{1}{2}(\overline{x}-\overline{x}\left(t;\overline{x}_0\right))^{\dagger} \overline{\overline{W}}(t) (\overline{x}-\overline{x}\left(t;\overline{x}_0\right))), \nonumber
 \label{eq:GAUSSIAN_TIME}
\end{equation}
where 
\begin{equation}
\overline{\overline{W}}(t) = \left(\overline{\overline{D}}(t)^{-1}\right)^{\dagger} \overline{\overline{W}}_0 \left(\overline{\overline{D}}(t)^{-1}\right). \nonumber
 \label{eq:DISPERIONS_TIME}
\end{equation}
The configuration-space density around the reference particle at
$\underline{x}(t;\overline{x}_0)$ is simply the integral of the
phase-space density over all velocities evaluated at
$\underline{x}(t;\overline{x}_0)$:
\begin{equation}
\rho(t) = \rho_0 \frac{\sigma_1(t)\sigma_2(t)\sigma_3(t)}{\sigma_0^3},
\end{equation}
where the velocity dispersions $\sigma_i(t)$ are given by
$1/\sqrt{\lambda_i(t)}$ and $\lambda_i(t)$ are the eigenvalues of the
velocity submatrix of $\overline{\overline{W}}(t)$.  The velocity
distribution in the principal axis frame of the velocity ellipsoid
centred on the particle's position and velocity is given by:
\begin{equation}
g\left(\underline{v}\right) = N(t) 
\exp(-\frac{1}{2} \underline{v}^\dagger \mathrm{diag}\left(\sigma_1(t),\,\sigma_2(t),\,\sigma_3(t)\right)^{-2} \underline{v}), \nonumber
\end{equation}
where $N(t)=1/((2\pi)^{3/2} \sigma_1(t)\sigma_2(t)\sigma_3(t))$. Note
that this velocity distribution is normalised, i.e. its integral over
velocity space is unity.

These quantities allow us to calculate the instantaneous annihilation
rate at each point on the particle's trajectory:
\begin{equation}
\frac{\mathrm{d}A}{\mathrm{d}t} = \frac{\rho(t)}{m} \int\mathrm{d}^3v
\, \sigma_A(v) v g\left(\underline{v}\right) =
\frac{\rho(t)\langle\sigma_Av\rangle}{m}
\label{eq:ANNIHILATION_RATE}
\end{equation}
where $m$ is the particle mass and $\sigma_A(v)$ the annihilation
cross-section.  We note that in many WIMP models the annihilation
cross-section is inversely proportional to encounter velocity, and in
this case $\langle\sigma_Av\rangle$ is independent of the actual local
velocity distribution \citep{1996PhR...267..195J}. An image of the system in annihilation
radiation can be constructed by integrating all particles forward over
a short time interval and summing their annihilation contributions
into a pixel array.  We note that equation (\ref{eq:ANNIHILATION_RATE})
exhibits near-singular behaviour as particles pass through caustics
and as a result special care is needed to obtain the correct
contribution to the annihilation luminosity in this situation.  This
will be discussed more fully in later papers.

\section{The DaMaFlow code} \label{sect:code}
We have developed the code DaMaFlow to test our GDE scheme by
analysing the behaviour of streams in a broad range of static
potentials. DaMaFlow integrates the equations of motion and in parallel
the GDE for a single orbit within user-specified potentials. This
requires solving $3+18$ second-order differential equations in
parallel.  The integration algorithm can be chosen to be a symplectic
second-order leapfrog (Drift-Kick-Drift or Kick-Drift-Kick
formulation) or alternatively the energy-conserving and adaptive
Dop853 algorithm \citep{HairerNorsettWanner1997/3} of order $8$ that
allows dense output and is very fast. Studies focusing on complex
phase-space structures, especially in the field of chaos analysis,
often use the Dop853 algorithm (or even higher order schemes) because
of its high precision.  On the other hand N-body codes often implement
the leapfrog method because it is the best compromise between
performance and accuracy. We find that with a moderate fixed
time-step, both formulations of leapfrog are able to give comparable
results to Dop853. This is an important point because it is not
possible to run N-body simulations with slow but accurate high-order
ODE solvers like Dop853.  

DaMaFlow is also set up to do a 
Numerical Analysis of Fundamental Frequencies (NAFF)
\citep{1988A&A...198..341L,1990Icar...88..266L,1996A&A...307..427P,
2003math......5364L} and of the
resonances associated with the chosen orbit. The fundamental frequencies are
revealed by an integer programming routine. This is needed so that
we can study the relation between the existence of well-defined
fundamental frequencies or resonances and stream density behaviour.
The NAFF method determines a quasi-periodic approximation to the
orbital motion. For ordinary Fast-Fourier-Transforms (FFT) the
accuracy of the determination of the frequencies is of the order of
$1/T$, where $T$ is the sampling interval. The NAFF method has an
accuracy of $1/T^4$. Thus it makes spectral analysis a lot faster
compared to classical Fourier techniques, for example, those used in
\cite{1982ApJ...252..308B}.

To scan large parts of phase-space, DaMaFlow can be run in parallel on
batch systems in order to integrate a large number of different orbits
simultaneously. A fast automated stream density fitting routine was
built in to facilitate efficient analysis of the underlying
phase-space without user interaction. Before this fitting can be done,
the stream density has to be smoothed to remove the large density
spikes produced by caustics.  DaMaFlow does this by extracting and
fitting the lower envelope of the stream density time series. This
envelope is constructed while the orbit integration is running by an
iterative on-the-fly minimum finder.

From a numerical point of view, solving the GDE is quite difficult in
chaotic regions of phase-space. In these regions the infinitesimal
phase-space volume around the reference particle gets distorted very
strongly. This produces large numerical values in the phase-space
distortion tensor. And this can lead to overflows and round-off errors
in numerical computations.  In chaos analysis it is an established
method to do some kind of re-norming to suppress these problems. We
can do something similar to follow the evolution of phase-space density
evolution. We use the transitivity of the phase-space
distortion tensor:
\begin{equation}
\overline{\overline{D}}_{t_1 \rightarrow t_3} = \overline{\overline{D}}_{t_2 \rightarrow t_3} \quad \overline{\overline{D}}_{t_1 \rightarrow t_2} 
\end{equation}
where $\overline{\overline{D}}_{t_i \rightarrow t_j}$ with $i<j$ is the solution of the GDE with 
$\overline{\overline{D}}_{t_i \rightarrow t_j}(t_i)=\overline{\overline{1}}$ evaluated at time $t_j$.
So the phase-space volume can be written as:
\begin{equation}
\mathrm{det}\left(\overline{\overline{D}}_{t_1 \rightarrow t_3}\right) = 
\mathrm{det}\left(\overline{\overline{D}}_{t_2 \rightarrow t_3} \right) \quad 
\mathrm{det}\left(\overline{\overline{D}}_{t_1 \rightarrow t_2} \right) 
\end{equation}
Thus dividing the time integration interval and re-initialising the
distortion after each interval avoids large numerical values. This is
very similar to the re-norming techniques used for calculating the
largest Lyapunov exponents of chaotic systems, where the re-norming
frequency is chosen to be of the order of the dynamical time-scale
\citep{0387907076,2002MNRAS.331...23E}.  Although this approach works
nicely for such applications, it does not help us when calculating the
stream density evolution, because we need the entire phase-space
distortion information from initial to final time. It is not possible
to separate the configuration-space part of the phase-space distortion
and to do a similar re-norming.  Thus one cannot avoid large numbers
during the calculation. DaMaFlow therefore calculates all quantities
in double-precision ($64$~bits=$8$~bytes).  Even in chaotic regions
this is enough to follow the system for a substantial amount of time.
We note that the phase-space density calculation involves the
determinant of a $6\times 6$ matrix, whereas the stream density only
involves the determinant of a $3\times 3$ matrix. As a result stream
density calculations are less strongly affected by large numbers and
overflows.  We note that special software libraries can provide even
higher precision (e.g. the GMP library
\footnote{\texttt{http://www.gmplib.org}}).  

Since we wish to implement the GDE formalism also into an N-body
code, execution speed and memory consumption are important
considerations. For each particle we need to store the full 6D
phase-space distortion tensor and the particle's tidal tensor. This
results in $36+6$ extra numbers per particle. Thus a $500^3$ particle
simulation in double-precision needs already about $39$~Gbytes of
random access memory (RAM) just for the GDE calculation, assuming we
store all information for every particle. 

\section{Integrable potentials} \label{sect:integrable}

\begin{figure}
\centerline{\includegraphics[width=1\linewidth]{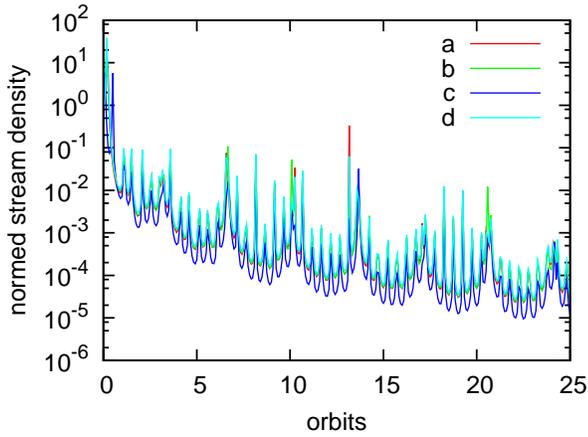}}
\caption{
\label{fig:SHEET_DENSITY}
Stream density evolution for different initial sheet orientations
$\underline{\underline{V}}_x (\overline{x}_0)$ in a static Hernquist
potential. The general shape of the four curves is very similar. The
long-term behaviour does not in general depend on the initial sheet
orientation.  }
\end{figure}

Before presenting some results for the evolution of stream densities
in integrable potentials we want briefly to discuss the choice of the
initial sheet orientation that goes into $\underline{\underline{V}}_x
(\overline{x}_0)$ in equation (\ref{eq:IC_CDM}).  This depends, of
course, on the starting time $t_0$ and the problem that is to be
studied.  For example, in a cosmological context
$\underline{\underline{V}}_x (\overline{x}_0)$ is given by the linear
initial conditions, so by the coupled initial density and velocity
fields.  We note that, to zeroth order, the sheet
orientation is simply given by the Hubble Flow:
$\underline{v}(\underline{x}_0,t_0)=H(t_0) \underline{x}_0$.

What is the impact of the initial orientation on later evolution?  In
Fig. \ref{fig:SHEET_DENSITY} we show the evolution of the normed
stream density, as calculated from eq. (\ref{eq:NORMED_FLOW_DENSITY}),
for a single orbit with four different choices of initial stream
orientation.  For this test we have used a spherical Hernquist
potential:
\begin{equation}
\Phi(r) =  - \frac{G M}{r+a}
\label{eq:HERNQUIST_POTENTIAL},
\end{equation}
with $M=1.86 \times 10^{12} \mathrm{M}_{\odot}$ and $a=34.5$ kpc. The reference particle begins its
orbit with $\underline{x}_0=(35,17.6,4)$ kpc and $\underline{v}_0=(-316.9,
-16.3, -4)$ $\mathrm{km}~\mathrm{s}^{-1}$. The initial sheet orientations were chosen to be 
(in units of $\mathrm{km}~\mathrm{s}^{-1}~\mathrm{kpc}^{-1}$):
$a:\underline{\underline{0}}$, $\,b:\underline{\underline{1}}$,
\begin{equation}
c: \left(
   \begin{array}{ccc}
    1 & 10 & 0 \\
    10 & 1 & 0 \\
    0 & 0 & 1 \\        
   \end{array}
\right)
\quad \mathrm{and} \quad
d: \left(
   \begin{array}{ccc}
    1 & 1 & -2 \\
    -1 & 1 & -1 \\
    2 & -1 & 1 \\        
   \end{array}
\right). \nonumber
\end{equation}

It is striking that all four curves have very similar shape and very
similar caustic spacings, although the caustic locations vary. The
long-term behaviour does not depend on initial sheet orientation,
at least in this case. Orientation $c$ produces lower densities than
the others because the scale of $\underline{\underline{V}}_x
(\overline{x}_0)$ is larger, but the shape of the lower envelope is
very similar. 

\cite{2006PhRvD..73b3510N} show that the caustic shape in
configuration-space is, in general, affected by the relative size of
the $\underline{\underline{V}}_x (\overline{x}_0)$ matrix elements.  A
detailed analysis of caustic shape thus requires choosing
$\underline{\underline{V}}_x (\overline{x}_0)$. For example, in their
model for the Milky Way halo \cite{2006PhRvD..73b3510N} initialised
trajectories at the turnaround sphere with a
$\underline{\underline{V}}_x (\overline{x}_0)$ loosely motivated by
tidal torque theory.  This restricted the form and scale of the
matrix, but still left a lot of freedom.  Here our main motivation is
not to analyse caustic shapes, but rather the long-term behaviour of
the fine-grained phase-space distribution, in particular of stream
densities. The initial sheet orientation is thus not an important
issue for us.  In the following we will consider quite general orbits,
but will arbitrarily set $\underline{\underline{V}}_x
(\overline{x}_0)=\underline{\underline{0}}$ unless otherwise stated
\footnote{Orbits starting on axes of symmetry in phase-space can show
non-generic stream density behaviour for $\underline{\underline{V}}_x
(\overline{x}_0)=\underline{\underline{0}}$ as we will discuss in the
section on non-integrable potentials (section 5).}.  Note that the
choice of $\underline{\underline{V}}_x (\overline{x}_0)$ does not
influence the dynamical evolution of the distortion tensor as it is
not part of the initial conditions for the GDE.  Only the final
projection to configuration-space is affected by initial sheet
orientation.

From a dynamical point of view, static, integrable potentials are very
simple. The motion within them can be described in terms of
action/angle variables and their Hamiltonian can be expressed solely
as a function of the actions $H=H(\mathbf{J})$. All motion in these
potentials is regular, so there are no chaotic regions in their
phase-space. In action-angle space the orbits lie on tori and are
characterised by a fixed number of fundamental frequencies.  DM
particles in integrable potentials will experience only phase mixing.

Because of these simple properties \cite{1999MNRAS.307..495H},
hereafter HW, were able to develop an analytic linearised treatment
based on action-angle variables to derive results for the stream
density behaviour. In their paper they did not specifically focus on
CDM, but rather analysed how Gaussian clouds in action-angle space
evolve with time.

As an example of an integrable potential, we apply our method to
several Eddington potentials $\Phi(r,\theta) = \Phi_1(r) + \eta(\beta
\cos{\theta})/r^2$ \citep{1962MNRAS.124...95L}. These are separable in
spherical coordinates.  The third integral for this type of potential
is $I_3 = \frac{1}{2} L^2 + \eta(\beta \cos{\theta})$. We chose the
following example of an Eddington potential,
\begin{equation}
\Phi(r, \theta) = v_{\rm h}^2 \log{(r^2 + d^2)} + 
\frac{\beta^2 \cos^2\theta}{r^2}, 
\label{eq:EDDINGTON_POTENTIAL}
\end{equation}
with $v_h=1$ $\mathrm{km}~\mathrm{s}^{-1}$, $d=1$ kpc, 
$\beta=2.5$ $\mathrm{kpc}~\mathrm{km}~\mathrm{s}^{-1} $
\footnote{These values do not have any specific meaning. We have chosen
them just in a convenient way.}, 
and studied an orbit which starts at
$\underline{x}_0=(5,3,2)$ kpc with a velocity of
$\underline{v}_0=(0.62,0.62,0.104)$ $\mathrm{km}~\mathrm{s}^{-1}$. 

\begin{figure}
\centerline{\includegraphics[width=1\linewidth]{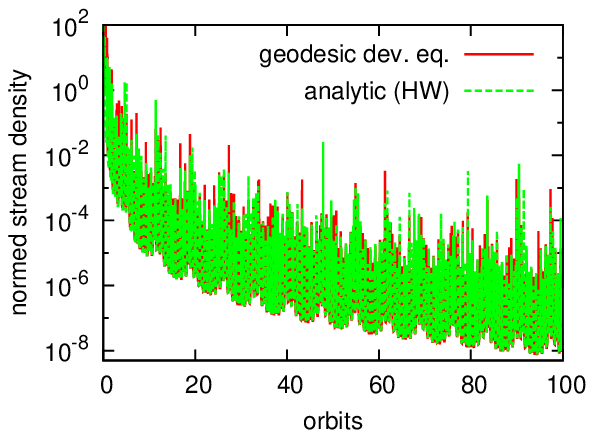}}
\centerline{\includegraphics[width=1\linewidth]{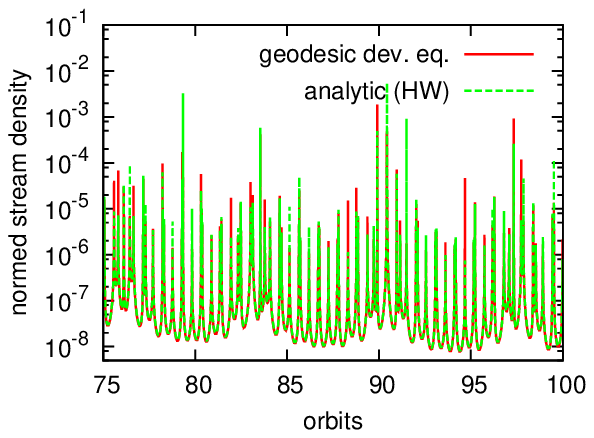}}
\caption{
\label{fig:EDDINGTON_DISTORTION}
The stream density evolution calculated using the GDE integrator
DaMaFlow is compared to the analytic result obtained by a linearised
treatment in action/angle variables \protect\citep{1999MNRAS.307..495H}. The
results agree essentially perfectly.  Notice how well the numerical
calculation matches the caustics.  The upper panel clearly shows that
the numerical result also has the correct
$1/(t/t_{\mathrm{orbital}})^3$ behaviour at late times. The initial
quasi-exponential decay is also visible.  }
\end{figure}

In Fig. \ref{fig:EDDINGTON_DISTORTION} we show the evolution of the
stream density for this orbit, i.e. the projection from phase-space to
configuration-space for an initial condition with
$\underline{\underline{V}}_x
(\overline{x}_0)=\underline{\underline{0}}$. Here and elsewhere
(unless otherwise stated) we define `the `orbital period'' as the
radial oscillation period for the purpose of making such plots. The
late-time behaviour of the stream density can be fitted by
an analytic formula derived by HW:
\begin{equation}
\rho(\underline{x},t) =  A \frac{1}{r^2 \sin\theta |p_r p_\theta|}\frac{1}{(t/t_{\mathrm{orbital}})^3},
\label{AXIS_LATE_FLOW}
\end{equation}
with only one fitting parameter $A$. Comparing this to the result for
the long-term behaviour in HW (Eq. 37) it is clear that $A$ just
reflects the initial phase-space distribution and the orbital
parameters (the derivatives of the fundamental frequencies with
respect to the actions). The results in HW are calculated for a
Gaussian cloud in phase-space, not a cold sheet as in our case.  Since
the initial distribution only affects $A$, the long-term behaviour of
the two configurations is the same, as shown in
Fig. \ref{fig:EDDINGTON_DISTORTION}.

One can clearly see that our method produces caustics at the correct
positions and is also able to recover the secular evolution of the
stream density, the $1/(t/t_{\mathrm{orbital}})^3$ density
decrease. We note also the initial quasi-exponential stream density
decrease that is often referred to as Miller's instability
\citep{1964ApJ...140..250M}. Recently \cite{2007arXiv0710.0514H}, hereafter HG, showed
this is a generic feature of Hamiltonian dynamics and not, as long
believed, an artifact specific to N-body integrations
\citep{2002ApJ...580..606H,2003ApJ...585..244K}.  HG discuss the
effect in detail for spherical potentials. All our tests with
spherical, axisymmetric and triaxial potentials show a similar
quasi-exponential initial decay. Since DaMaFlow integrates the
equations of motion for a single particle in a perfectly smooth
potential, it is evident that this behaviour has nothing to do with
N-body effects.

\begin{figure}
\centerline{\includegraphics[width=1\linewidth]{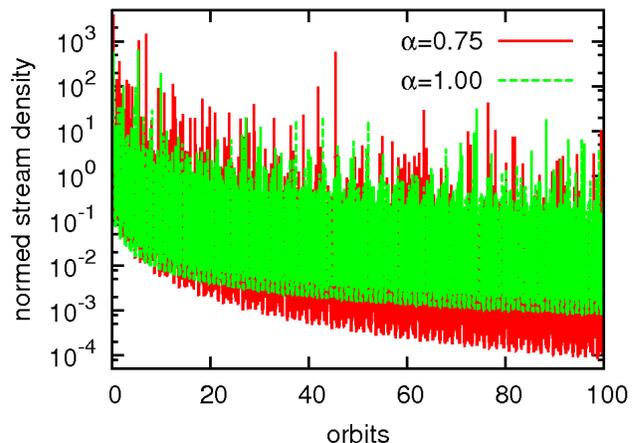}}
\caption{
\label{fig:MOD_KEPLER}
Stream density evolution for the normal Kepler potential ($\alpha=1$)
and for a modified potential ($\alpha=0.75$). Orbits in the Kepler
potential have a single fundamental frequency. Changing the potential
exponent from $\alpha=1$ to $\alpha=0.75$ increases the number of
fundamental frequencies to two. As a result, the long-term stream
density behaviour changes from $1/(t/t_{\mathrm{orbital}})$ to
$1/(t/t_{\mathrm{orbital}})^2$. For integrable potentials the
long-term stream density decrease on each orbit is dictated by the
number of fundamental frequencies.}
\end{figure}

A NAFF frequency analysis of particle orbits in the
Eddington potential reveals, as expected, three linearly independent
frequencies. It is this number that dictates the speed with which
stream densities decrease in static, separable potentials. One can see
this very clearly from Fig. \ref{fig:MOD_KEPLER}. Here we show the
density decrease in a simple Kepler-like toy-model:
\begin{equation}
\Phi(r) =  - \frac{1}{r^\alpha}.
\label{eq:KEPLER}
\end{equation}
For $\alpha=1$ the orbit was started from $\underline{x}_0=(-0.33,
0.97, 0)$ and $\underline{v}_0=(-1.0, -0.07, 0)$ with an energy
$E=-0.5$.  This orbit has only one fundamental frequency.  Changing
$\alpha$ to $0.75$ increases the number of frequencies to two; the
loops of the orbit no longer close. The starting point for this second
case has been set to $\underline{x}_0=(-0.27, 1.28, 0)$ and
$\underline{v}_0=(-0.76, 0.24, 0)$, corresponding also to $E=-0.5$.
The increased number of frequencies results in a more rapid decrease
in stream density: $1/(t/t_{\mathrm{orbital}})$ for one fundamental
frequency, and $1/(t/t_{\mathrm{orbital}})^2$ for two. In this sense
the long-term behaviour of streams in static, integrable systems is
very simple and is determined only by the number of fundamental
frequencies.

\section{Non-integrable potentials} \label{sect:nonintegrable}

Analytic methods like the HW formalism are only able to deal with
integrable potentials.  This is a serious limitation since realistic
potentials are rarely integrable. To demonstrate that the GDE method
can also deal with more complex phase-space structure we now discuss
the well-known ellipsoidal logarithmic potential with a core,
\begin{equation}
\Phi(x,y,z) = \frac{1}{2} v_0^2 \ln \left(r_c^2 + x^2 + \left(y/q\right)^2 + \left(z/p\right)^2\right),
\label{eq:LOG_POTENTIAL}
\end{equation}
analysing its stream density behaviour and its phase-space
structure. There are two reasons why we have chosen this potential:
first there has been substantial previous work on its phase-space
structure \citep{1982ApJ...252..308B, 1998A&A...329..451P}, so we can compare directly with these earlier
results. Second this potential is often considered as a good model for
galactic halos because it reproduces a flat rotation curve and its
shape can easily be tuned by two parameters that correspond to the
axial ratios of the ellipsoidal isopotential surfaces. It has been
used, for example, for dynamical studies of the debris streams of the
Sagittarius dwarf galaxy \citep{2004MNRAS.351..643H}. The ellipsoidal
logarithmic potential without a core ($r_c=0$) has also been used to
study the influence of halo shape on the annual modulation signal in
dark matter detectors \citep{2000MNRAS.318.1131E}.

\begin{figure}
\centerline{\includegraphics[width=1\linewidth]{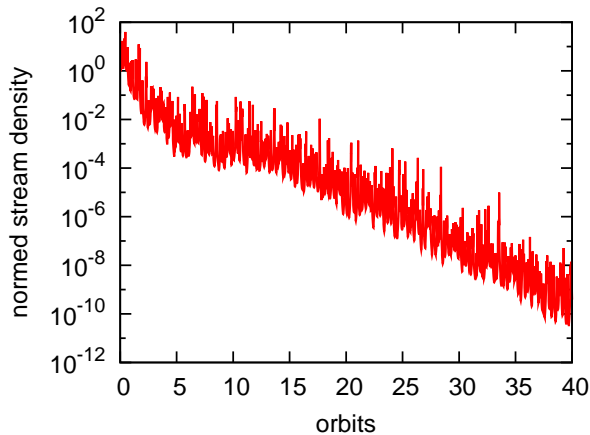}}
\caption{
\label{fig:LOG_CHAOTIC_FLOW}
Stream density evolution along a chaotic orbit in the ellipsoidal
logarithmic potential with $q=1.5, p=0.5$. The density decreases very
rapidly, reflecting the chaotic mixing along this orbit. Note that the
decrease is no longer a power-law, as in the case of regular motion.
}
\end{figure}

It is known that, depending on the degree of triaxiality, the
phase-space of the logarithmic potential can be occupied to a large
extent by chaotic orbits \citep{1998A&A...329..451P}. In
Fig. \ref{fig:LOG_CHAOTIC_FLOW} we show how the stream density evolves
along one of these chaotic orbits.  This orbit was integrated in a
potential with $q=1.5, p=0.5, v_0=1, r_c=1, E=3$ and started at
$\underline{x}_0=(10,1,5)$,$\,\underline{v}_0=(0.16,80,-0.16)$.
Here we apply the same system of units as \cite{1998A&A...329..451P} 
and write all quantities as dimensionless numbers.
It is evident from this plot that the system mixes very rapidly along this
orbit. This is chaotic mixing, and contrasts with the phase mixing
that we found before for regular motion in separable potentials.  We
note that chaotic orbits are difficult to handle from a numerical
point of view because of the rapid spread in phase-space that
characterises them. To check whether we can rely on our stream density
values, we also calculated the 6-D phase-space density along this
orbit. Over the full integration range shown in
Fig. \ref{fig:LOG_CHAOTIC_FLOW} ($40$ orbits) it remained constant to
an accuracy of $10^{-7}$.

Using DaMaFlow we have scanned the phase-space of box orbits within
the logarithmic potential, identifying chaotic and regular regions by
calculating the stream density evolution for about $5 \times 10^4$
different orbits. Each orbit was integrated for a fixed time interval
of $2000$ using $2\times 10^5$ time-steps. This corresponds to about
$10^3$ orbital periods. We chose this very long integration time in
order to distinguish between chaotic and regular
regions\footnote{For chaotic orbits elements of the distortion
matrices can become very large. Here we are only interested in
separating chaotic and regular orbits reliably.  Thus we do not care
about round-off errors in the calculation of these matrices.}.

\begin{figure}
\centerline{\includegraphics[width=1.2\linewidth]{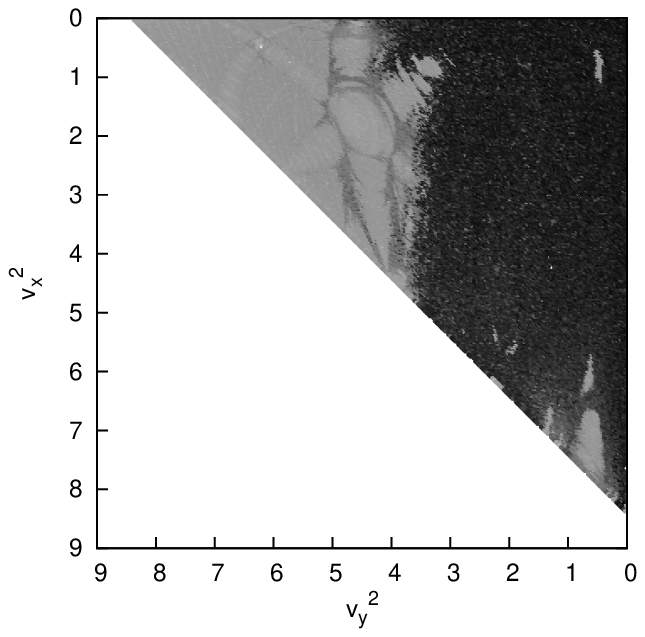}}
\centerline{\includegraphics[width=1.2\linewidth]{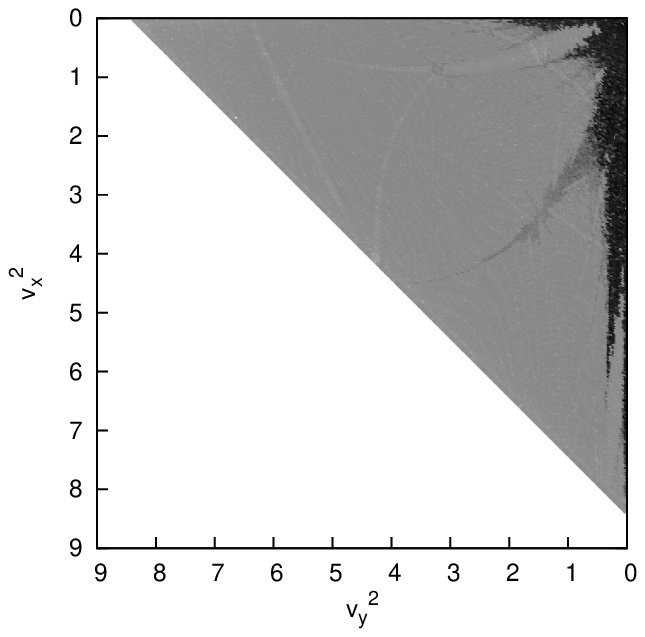}}
\caption{
\label{fig:LOG_CHAOTIC_MAP}
Chaos maps for box orbits in the logarithmic ellipsoidal potential for
two different sets of axial ratios. The upper plot is for a highly
aspherical potential with $q=1.8, p=0.9$, whereas the lower plot is
for a rounder potential with $q=1.1, p=0.9$.  The greyscale indicates
the stream density decrease after a fixed time interval (2000 time
units or about 100 orbital periods). Black regions mark the very low
final stream densities found for chaotic orbits, while grey regions
mark the higher stream densities found for regular orbits. Densities
decay quasi-exponentially in the former case, but only as a power law
of time in the latter.  These plots can be directly compared to
Fig.~6(c) and Fig.~4(b) in \protect\cite{1998A&A...329..451P}, where a
frequency shift analysis of the same system reveals exactly the same
structures. }
\end{figure}

Our results can be compared directly to previous work where the same
potential was analysed using frequency shifts
\citep{1998A&A...329..451P}.  This method is based on the fact that
chaotic motion, contrary to regular motion, has no stable fundamental
frequencies, so that frequency estimates shift with time. By looking
for such shifts one can distinguish between chaotic and regular
motion.  Fig. \ref{fig:LOG_CHAOTIC_MAP} shows maps of orbit type for
two different sets of axial ratios $q, p$. This figure can be compared
directly to Fig.~6(c) and Fig.~4(b) in \cite{1998A&A...329..451P}. For
the first of these calculations we adopted the following values:
$q=1.8, p=0.9, v_0=\sqrt{2}, r_c=0.1$ and $E=-0.404858$. For the
second case we changed the potential shape by instead taking $q=1.1,
p=0.9$. 

The maps of Fig. \ref{fig:LOG_CHAOTIC_MAP} are constructed as follows.
We start each orbit at the centre of the potential to be sure to get a
box-orbit with zero angular momentum. Each individual orbit can 
be labelled by its initial $v_x$ and $v_y$ velocity components.  The
value of $v_z$ is then determined by the chosen value of the energy,
$E=-0.404858$. This is, of course, a very special point within the
potential, and it turned out that our standard choice of initial
stream orientation, $\underline{\underline{V}}_x
(\overline{x}_0)=\underline{\underline{0}}$, produces highly
non-generic stream density behaviour in this case; the stream density
remains constant! In order to get properly representative behaviour we
therefore took $\underline{\underline{V}}_x
(\overline{x}_0)=\underline{\underline{1}}$ when producing
Fig. \ref{fig:LOG_CHAOTIC_MAP}.  With this set-up, we scanned the
whole $v_x - v_y$ plane and saved the stream density of each orbit
after a fixed amount of time.  The greyscale in the plots corresponds
to the stream density decrease after that fixed time. Black points
denote a very rapid stream density decrease, thus regions of chaotic
mixing. Grey regions show a slower density decrease, reflecting phase
mixing and regular motion. Much of the box-orbit phase-space is
chaotic for $q=1.8, p=0.9$. Reducing the asphericity to $q=1.1, p=0.9$
results in a much larger fraction of the boxes being regular.

A comparison of Fig.~\ref{fig:LOG_CHAOTIC_MAP} to Fig.~6(c) and
Fig.~4(b) in \cite{1998A&A...329..451P} shows excellent and detailed
agreement.  Regions of high frequency shift correspond, as expected,
to those of rapid stream density decrease and chaotic mixing. Thus the
GDE and the frequency shift technique work equally well for
delineating regions of chaotic and normal phase mixing. We note that
the Lyapunov exponent technique for identifying chaotic behaviour is
closely related to the GDE \citep{0387907076, 2002MNRAS.331...23E},
since these exponents are obtained from the eigenvalues of the 6D
distortion tensor.  Identifying and characterising chaotic behaviour
is not the main goal of our work here, so we will not pursue this
connection further in this paper.

So far we have classified orbits as either regular or chaotic, but the
regular part of phase-space has substructure in the form of resonances
\citep{1998MNRAS.298....1C, 1998MNRAS.298...22W, 1999AJ....118.1177M}.
These are regions where the frequencies of motion are commensurate
$m_1 \omega_1 + m_2 \omega_2 + m_3 \omega_3$ where the $m_i$ are
integers and the $\omega_i$ are the three frequencies of the regular
motion. As shown above, resonance influence the stream density
behaviour over long timescales
\citep{1999MNRAS.307..495H,2007arXiv0710.0385S}. This is because they
restrict the motion to a lower dimensional region in phase-space,
implying that the orbit does not fill its KAM
(Kolmogorov-Arnold-Moser) torus densely. In simple terms, the system
cannot spread as fast as it would do in the non-resonant case because
it occupies a space of lower dimension.

\begin{figure}
\centerline{\includegraphics[width=1\linewidth]{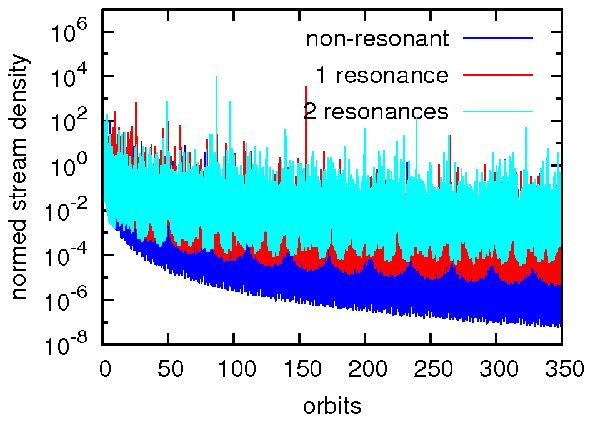}}
\centerline{\includegraphics[width=1.1\linewidth]{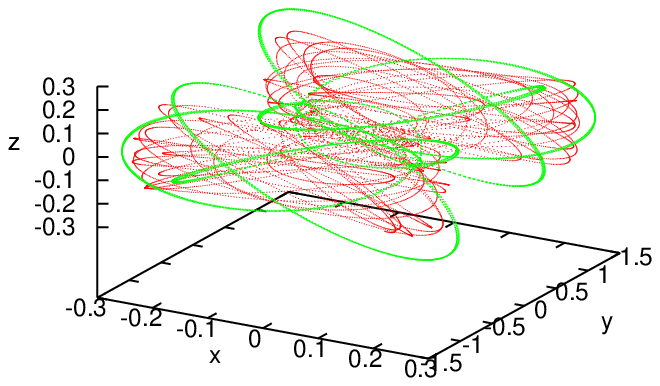}}
\caption{
\label{fig:RESONANT_ORBITS}
Stream density evolution on resonant orbits.  Stream density drops at
a rate which depends on the number of independent orbital frequencies.
Non-resonant orbits have three independent frequencies, and their
stream density decreases like $(t/t_{\mathrm{orbital}})^{-3}$ at late
times.  Resonances reduce the number of independent frequencies and
slow the decrease of stream densities. With one resonance there are
two independent frequencies; the stream density then falls as the
inverse square of time. With two resonances, only one independent
frequency remains and density falls as the inverse first power of
time.  The number of resonances also strongly affects the orbit shape
in configuration-space.  As is visible for the examples in the lower
plot, non-resonant orbits (red) fill a 3D volume, whereas orbits with
two resonances (green) are restricted to a line.  }
\end{figure}

Fig.~\ref{fig:RESONANT_ORBITS} shows the stream density evolution
along three different box orbits for $q=1.8, p=0.9, v_0=\sqrt{2},
r_c=0.1, E=-0.404858$.  The initial conditions for these orbits were
chosen so that they have different numbers of orbital resonances (non-resonant,
one resonance, two resonances):
$\underline{x}_0=(-0.08,-0.70,-0.090)$, $\underline{v}_0=(-0.60,-1.50,-0.07)$  , 
$\underline{x}_0=(-0.01, -0.67, -0.08)$, $\underline{v}_0=(-0.53,-1.60, -0.06)$   and
$\underline{x}_0=(0.08,-0.63,-0.14)$, $\underline{v}_0=(-0.51,-1.51,-0.52)$. 
It is clear
that the number of resonances has a major effect on the final stream
density decrease over timescales similar to those shown in this
plot. The difference in stream density between the non-resonant case
and the case with two resonances is about 3 orders of magnitude after
350 orbits! Resonances also have a strong influence on the shape of
the orbit in configuration-space, as shown in the lower panel of
Fig.~\ref{fig:RESONANT_ORBITS}. Two resonances restrict the orbit to a
line. From the shape of the orbits it is evident why the stream
density changes so much with the resonances. The particles cannot
spread over a large region if they are confined to a space of small
dimension.

\begin{figure}
\centerline{\includegraphics[width=1.0\linewidth]{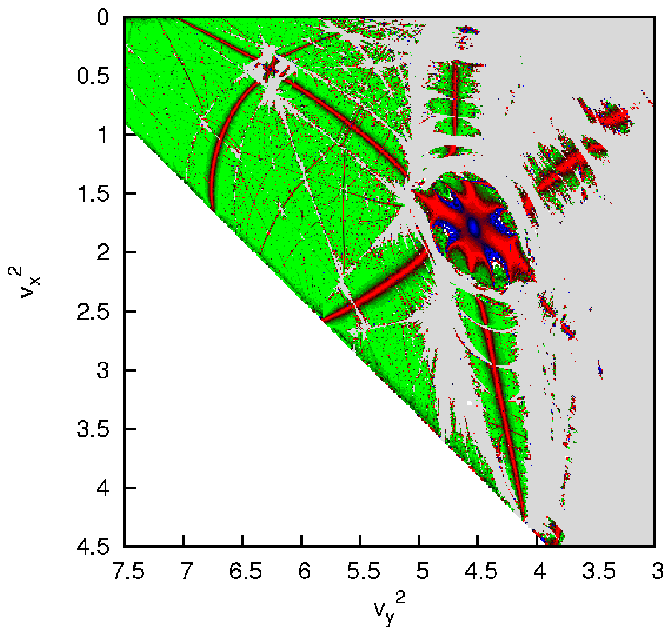}}
\caption{
\label{fig:RESONANT_SCAN}
Resonance structure of a logarithmic potential as revealed by the
GDE. Green indicates regions with no resonance (i.e. three independent
orbital frequencies). Stream densities for these orbits decrease
as $(t/t_{\mathrm{orbital}})^{-3}$ at late times.  Red indicates
orbits with one resonance, for which stream densities decreases as
$(t/t_{\mathrm{orbital}})^{-2}$. Orbits with two resonances are
coloured blue.  This map was constructed by integrating $450\times450$
orbits, each for about $10^4$ orbital periods, using DaMaFlow in
parallel.  Each orbit required $25\times10^6$ time-steps using a KDK
leapfrog algorithm.  }
\end{figure}

The regular region of phase-space for the logarithmic potential is
occupied by resonant and non-resonant regions. We used DaMaFlow to
scan the regular region (as for the chaos maps above) and fitted the
stream density decrease by a power law in time. A non-resonant motion
then gives an exponent of $3$, whereas regions with two resonances
should give $1$.  We binned these power law exponents (bin size 0.1)
and coloured them.  The result of this procedure is shown in
Fig. \ref{fig:RESONANT_SCAN}.  For this map we integrated a total
$450\times 450$ orbits for about $10^4$ orbital periods. Each 
integration required $25\times10^6$ KDK leapfrog time-steps.  The chaotic
regions are shown in grey, while the regular regions are shown in
three different colours depending on their stream density
behaviour. We note that the chaos detection here was carried out by imposing
a threshold $10^{-15}$ on the stream density decrease. Every orbit
that crosses this threshold during its evolution is considered as
chaotic and marked as grey in the map. This is the reason why the
chaotic pattern here is not identical to that in
Fig.~\ref{fig:LOG_CHAOTIC_MAP}.  

Most of the regular phase-space in this figure is occupied by
non-resonant orbits shown in green.  Superposed on these is a fine
network of resonance lines shown in red. In blue regions the stream
density decreases linearly with time because there are two resonances.
Note how well our method locates the resonance lines in the initial
condition space spanned by $v_y$ and $v_x$.
\cite{1998A&A...329..451P} analysed resonances with frequency maps by
plotting the rotation numbers defined as $a_1=\nu_L/\nu_S$ and
$a_2=\nu_M/\nu_S$, where $\nu_i$ are the fundamental frequencies along
the long (L), short (S) and middle (M) axes.  It is straightforward to
identify the resonance lines in Fig.~\ref{fig:RESONANT_SCAN} with
those in the frequency map of \cite{1998A&A...329..451P} by applying a
NAFF frequency analysis to the orbit corresponding to any specific
initial condition, for example, one on a given resonance line.  It
turns out that we can identify all resonance lines in
Fig. \ref{fig:RESONANT_SCAN} with similar lines in the frequency
map. At the intersection of these lines we have periodic orbits
(satisfying two resonance conditions) with the same rotation numbers
as those found in \cite{1998A&A...329..451P}. For example, the line
going from the upper left corner to the lower right corresponds to the
$(3,1,-2)$ resonance, meaning that $3 a_1 + 1 a_2 - 2 = 0$.

We conclude that our method can resolve the structure of phase-space
equally as well as the standard frequency mapping technique.

\section{Triaxial Dark Matter halos} \label{sect:triaxial}

\begin{figure}
\centerline{\includegraphics[width=0.9\linewidth]{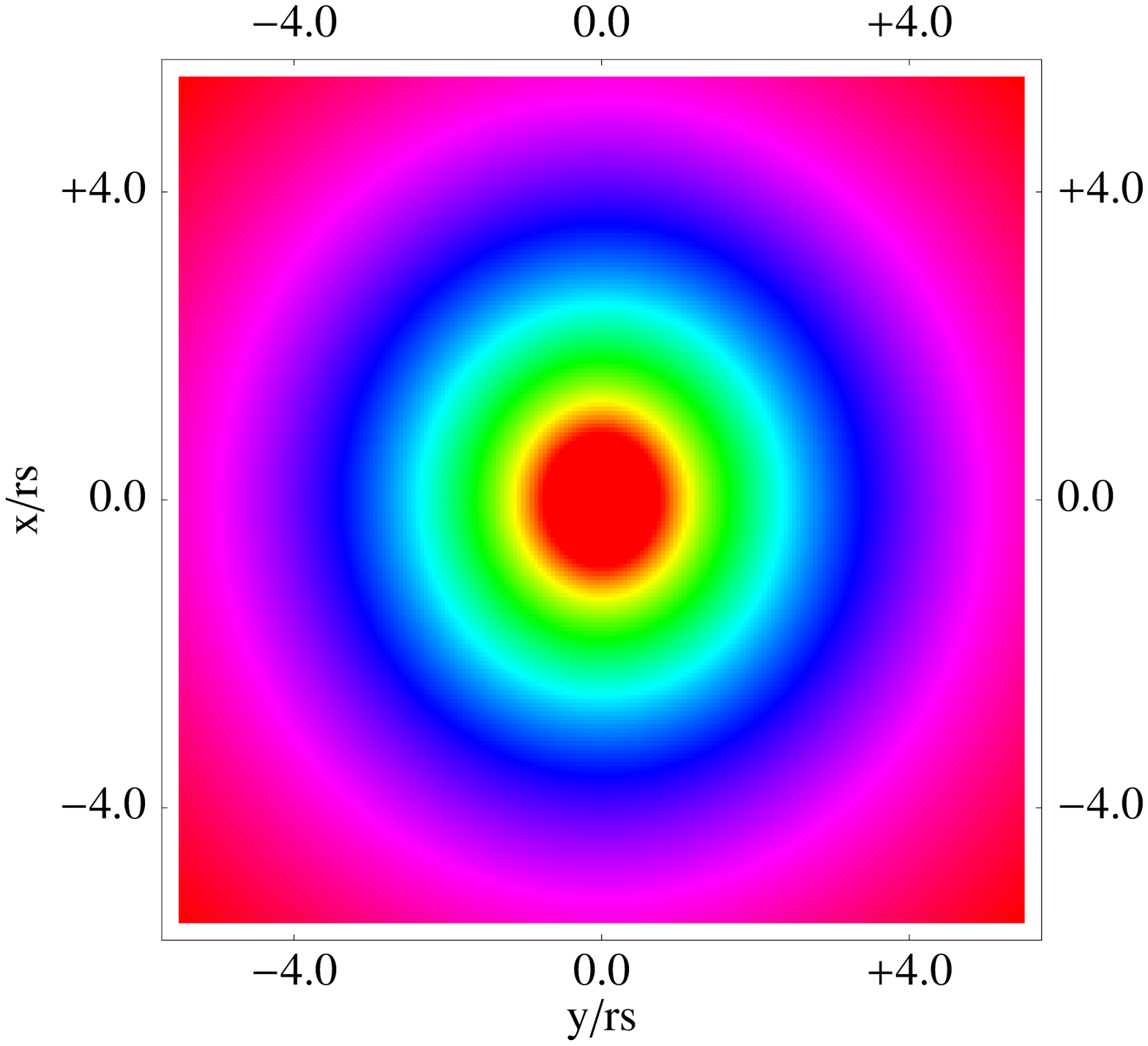}}
\centerline{\includegraphics[width=0.9\linewidth]{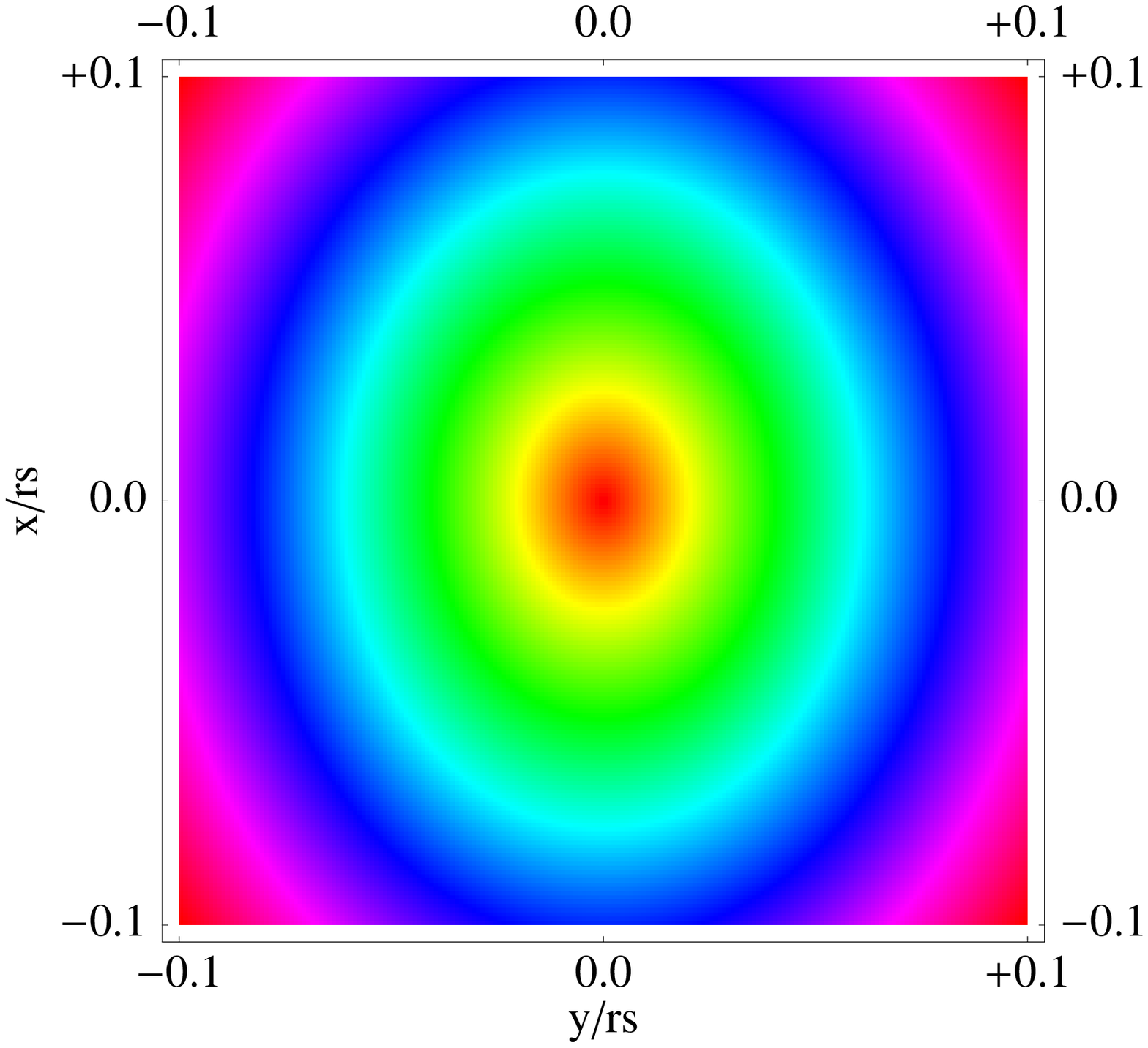}}
\caption{
\label{fig:ISOPOTENTIALS}
Isopotentials for the outer and inner parts of one of our triaxial NFW
halos.  It is obvious that the halo becomes rounder as one moves
outwards. In this case the transition scale $r_a$ was chosen to be
equal to the scale radius $r_s$ of the NFW profile.  }
\end{figure}

\begin{figure}
\centerline{\rotatebox{270}{\includegraphics[width=0.7\linewidth]{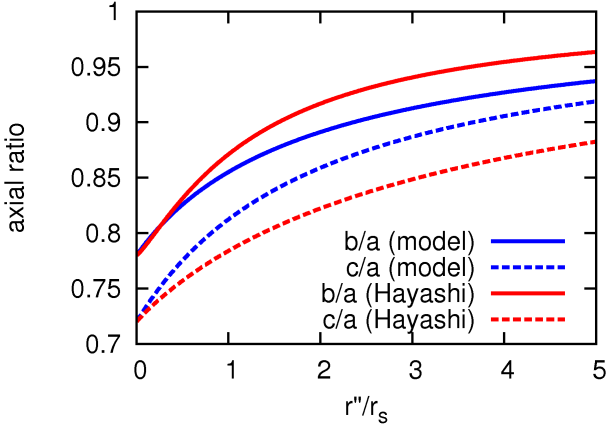}}}
\caption{
\label{fig:AXIAL_RATIOS}
Comparison of the radial variation of isopotential axial ratios for
N-body halos \protect\citep{2007MNRAS.377...50H} to that for our simple
triaxial NFW model.  The N-body values are the average of those found
in \protect\cite{2007MNRAS.377...50H}. The transition scale $r_a$ of our
model has been set equal to the scale radius $r_s$ of the underlying
NFW profile.  }
\end{figure}

In CDM cosmologies dark matter halos are not spherical. Furthermore,
simulations suggest that their shape should vary with radius, both
equidensity and equipotential surfaces being rounder (on average) at
larger radii. Several studies have tried to constrain the shape of the
Milky Way's halo by analysing the properties of observed tidal streams
like that of the Sagittarius dwarf galaxy
\citep{2001ApJ...551..294I,2004MNRAS.351..643H}.  Recently,
\cite{2007MNRAS.377...50H} analysed the radial variation in potential
shape of simulated halos that might correspond to that of the Milky
Way.  Although there is substantial object-to-object scatter, on
average they found a relatively rapid transition from aspherical to
almost spherical which occurs near the scale radius $r_s$ of the best
fitting NFW profile. They provide a simple fitting formula for this
mean behaviour,
\begin{equation}
\log \left({\frac{b}{a}} \ {\rm or}\ {\frac{c}{a}}\right) = \alpha \left[ \tanh \left(
  \gamma \log \frac{r}{r_{\alpha}} \right) -1 \right],
\label{eq:FIT_SHAPE}
\end{equation}
for the principal axial ratios $b/a$ and $c/a$. (Note that they
actually provide two different sets of fitting parameters for
Eq. (\ref{eq:FIT_SHAPE}) depending on the principal axial ratios.)
They also propose a modified NFW potential that takes into account the
variation in shape, but this potential is not very convenient because
it is not straightforward to derive the corresponding equations of
motion.  The examples given above show that potential shape can have a
substantial effect on stream density evolution, so it is interesting
to see how strong such effects can be in a realistic model.

To analyse this we have built a simple extension of the NFW model that
qualitatively reproduces the shape variation found by
\cite{2007MNRAS.377...50H} but which has simple equations of motion
that can easily be implemented in DaMaFlow. (For another similar
model, see \cite{2007arXiv0708.3101A}.)

We model the variable shape of the NFW halo by replacing the euclidean
radius in the formula for the potential of a spherical NFW halo by a
more general ``radius" $\tilde r$ given by:
\begin{equation}
 \tilde{r} = \frac{\left(r_a + r\right) r_E }{\left(r_a + r_E\right)}.
 \label{RADIUS_TILDE}
\end{equation}
Here $r_a$ is a transition scale where the potential shape changes
from ellipsoidal to near spherical and $r_E$ is an ellipsoidal
``radius'' given by:
\begin{equation}
 r_E = \sqrt{\left(\frac{x}{a}\right)^2 + \left(\frac{y}{b}\right)^2 + \left(\frac{z}{c}\right)^2},
 \label{eq:ELLIPSOIDAL_RADIUS}
\end{equation}
where we require $a^2 + b^2 + c^2 = 3$.  Thus for $r \ll r_a$ $\tilde
r \cong r_E$ and for $r \gg r_a$ $\tilde r \cong r$. We then take the
potential to be $\Phi(x,y,z) = \Phi_{\mathrm{NFW}}(\tilde r(x,y,z))$
which reproduces the general behaviour found by
\cite{2007MNRAS.377...50H} with a smooth transition around $r_a$.

For a specific example, we have chosen the transition scale to be
the scale radius of the NFW profile and have taken values for
$a, b$ and $c$ that give central principal axial ratios that are
comparable to those found by Hayashi et al: $b/a=0.78$ and $c/a=0.72$.
Our choice is $a=1.18,b=0.92,c=0.85$. For the NFW profile we used a
concentration of $r_{200}/r_s=7.0$. We checked Poisson's equation for
this potential to ensure that it implies a positive density
everywhere.  The check was performed by DaMaFlow evaluating the
negative of the trace of the tidal field on a fine 3D grid. This is just the
Laplacian of the potential and so proportional to the corresponding
density. Since the density field is continuous, positive density
values on the grid should guarantee a positive density everywhere.

\begin{figure*} 
\centerline{\rotatebox{0}{\includegraphics[width=1.0\textwidth]{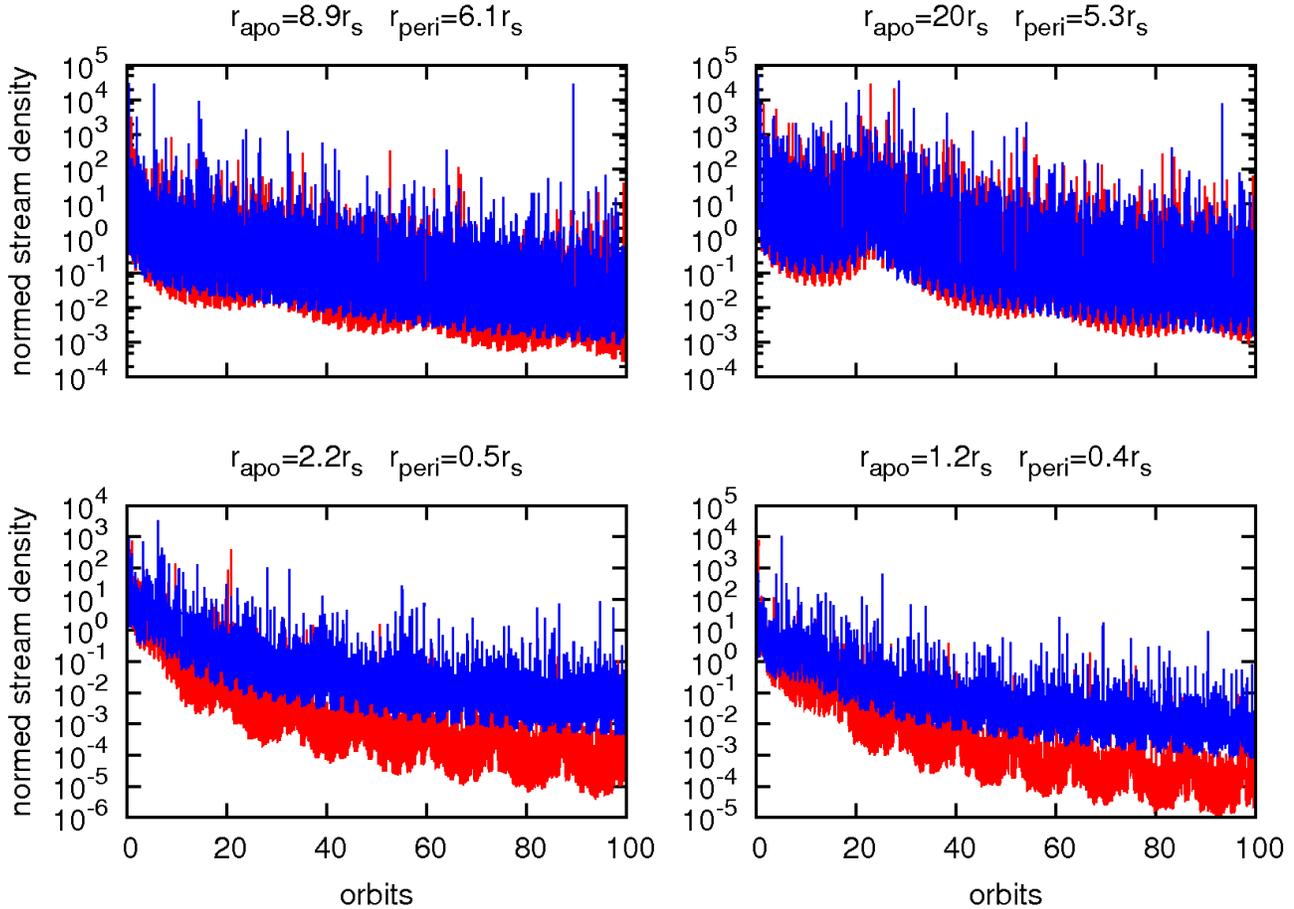}}}
\caption{ Stream densities in a spherically symmetric NFW potential
are compared to those expected in a more realistic DM halo with a
shape that varies with radius. Orbits with pericentre inside the
transition scale $r_a=r_s$ show a substantial difference between the
two cases. After about 100 orbits streams are roughly $100$ times less
dense in a triaxial halo than in a spherical one. Thus spherical
models for the Milky Way's halo are likely to underestimate the number
of streams in the solar neighbourhood by two orders of magnitude.}
\label{fig:TRIAXIAL_FLOWS} 
\end{figure*}

Fig.~\ref{fig:ISOPOTENTIALS} shows isopotentials in the outer and
inner parts of the halo. All distances are expressed in terms of the
scale radius $r_s$ of the NFW profile. The transition from spherical
to aspherical can clearly be seen as the centre is approached.
Fig.~\ref{fig:AXIAL_RATIOS} compares the radial variation in axial
ratios in our model and in the simulations of
\cite{2007MNRAS.377...50H}. The simulation axial ratios are calculated
with eq. (\ref{eq:FIT_SHAPE}) using the average values for $\alpha,
\gamma, r_\alpha$ found in \cite{2007MNRAS.377...50H}. The lines for
our model are calculated as follows. For a given value of $\tilde r$
we computed the intersections of the corresponding isocontour with the
$x, y$ and $z$ axes. So we get three values $a^{\prime \prime},
b^{\prime \prime}, c^{\prime \prime}$. To look for their variation
over distance we define the mean distance $r^{\prime \prime} =
\sqrt{(a^{\prime \prime})^2 + (b^{\prime \prime})^2 + (c^{\prime
\prime})^2}$. This is essentially the same procedure which
\cite{2007MNRAS.377...50H} applied when fitting the isopotentials of
their simulated halos. Thus we can compare directly with their results
as in Fig.~\ref{fig:AXIAL_RATIOS}. The qualitative behaviour of our
model is very similar to that of the simulations.  It is not necessary
to demand an exact fit since the scatter between different halos
studied by \cite{2007MNRAS.377...50H} is quite large.

We implemented this potential into DaMaFlow and looked at four
different orbits with the following apo-/pericentre distances in units
of $r_s$: $8.9/6.1$,$\,20/5.9$,$\,2.2/0.5$,$\,1.2/0.4$.  We compared
the stream densities predicted for our triaxial model to those
predicted in the corresponding spherical NFW profile. We fixed the
starting point and the velocity direction to be the same in the two
cases. The amplitude of the velocity was then set to give the same
energy in the two cases. With this procedure the orbits covered
comparable regions in configuration-space and had nearly the same apo-
and pericentre distances. Note that it is impossible to get identical
orbital shapes in the two potentials.  Especially in the inner parts
of the halo, where the two potentials differ substantially, the orbits
have different shapes. In a spherical potential orbits are confined to
a plane by conservation of angular momentum, but this is not the case
in a triaxial potential.

In Fig.~\ref{fig:TRIAXIAL_FLOWS} we show the stream density evolution
for these four orbits. Two belong to the outer halo with peri- and
apocentre beyond the scale radius. As expected their stream density
behaviour is very similar in the two potentials.  As soon as orbits
penetrate the inner halo, however, the behaviour is quite different in
the two cases. After $100$ orbits, streams are about 100 times less
dense in a triaxial halo than in a spherical one. Note that this is
just what one would predict, given the $(t/t_{\mathrm{orbital}})^{-2}$
and $(t/t_{\mathrm{orbital}})^{-3}$ density evolution expected for
regular motion in spherical and triaxial potentials, respectively.

In the case of a cored ellipsoidal potential we showed above that,
depending on the level of triaxiality, substantial fractions of
phase-space can be occupied by chaotic orbits. Thus we may expect such
orbits to be present in our triaxial NFW profile also.  Previous work
on galactic dynamics has demonstrated the presence of chaotic orbits
in the potentials corresponding to a variety of cuspy, triaxial
density profiles \citep{1996ApJ...460..136M, 1996ApJ...471...82M,
1998ApJ...506..686V,2003MNRAS.345..727K,2007ApJ...666..165C}.  To search for chaotic
orbits in our model, we have integrated $2 \times 10^4$ different
representative orbits and studied the predicted behaviour for
the density of their associated streams.  For simplicity we have
chosen the initial conditions for these orbits at random from the
known analytic distribution function of a Hernquist sphere matched to
the mean radial density profile of our triaxial NFW model. A
self-consistent Hernquist
sphere has density profile and potential:
\begin{equation}
\rho(r) = \frac{M}{2 \pi} \frac{a}{r} \frac{1}{(r+a)^3};~~~~\Phi(r) = -\frac{G~M}{r+a}.
 \label{eq:HERNQUIST_DENSITY}
\end{equation}
We match to our NFW model by an appropriate choice of the scale length
$a$ \citep{2005MNRAS.361..776S}. A standard inversion technique can
then be used to select a random set of initial orbital positions from
this (spherically symmetric) distribution. Appropriate initial
velocities can be generated by applying the von Neumann rejection
technique \citep{573140,2005MNRAS.356..872A} to the analytically known
distribution function of the self-consistent Hernquist model
\citep{1990ApJ...356..359H}. This procedure does not, of course,
sample orbits with a weighting which would self-consistently reproduce
our triaxial NFW model.  Nevertheless, the similarity of the NFW and
Hernquist models should ensure that our selected orbits cover the
regions of phase-space which would be populated in a truly
self-consistent model in a reasonably representative way.  This is
sufficient to evaluate the overall importance of chaotic orbits in the
model. 

\begin{figure}
\centerline{\rotatebox{0}{\includegraphics[width=1.2\linewidth]{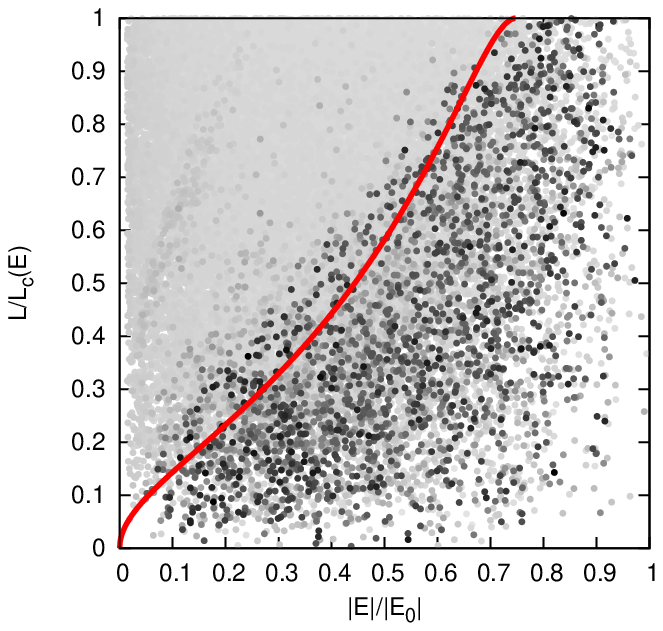}}}
\centerline{\rotatebox{0}{\includegraphics[width=1.2\linewidth]{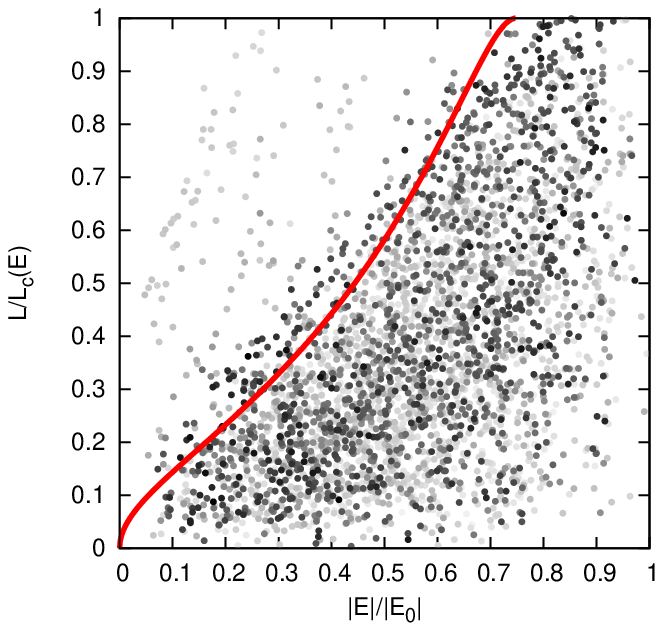}}}
\caption{
\label{fig:NFW_TRIAXIAL_CHAOS}
Qualitative view of the phase-space structure of our triaxial NFW
potential. These plots give results for $2 \times 10^4$ orbits. 
The greyscale represents the stream density decrease after 
$\mathrm{max}(x_\mathrm{cross},y_\mathrm{cross},z_\mathrm{cross}) = 10^3$.
Black points correspond to orbits with a very strong density decrease, thus to non-regular orbits.
Such orbits are found primarily in the inner regions of the potential.
Orbits with lower binding energy are mostly regular, showing a much
smaller stream density decrease.  The black points were plotted last
to avoid over-plotting by the grey ones. $E_0$ corresponds to a
particle at rest at the centre of the potential. The solid red line
represents the constant pericentre line ($r_\mathrm{peri}=1/3 r_s$).
In the lower panel only the box orbits are plotted. Most of the non-regular
orbits are boxes deep inside the potential.
}
\end{figure}

We integrated each orbit as long as required to guarantee
$\mathrm{max}(x_\mathrm{cross},y_\mathrm{cross},z_\mathrm{cross}) = 10^3$, 
where $i_\mathrm{cross}$ is the number of crossings along the $i$-coordinate,
$i=x,y,z$.
The integration was done by the Dop853 algorithm with a very high precision
to get an relative energy error below $10^{-10}$ over the whole integration time.
We have chosen such high energy conservation to be sure that the
integration works correctly even though the potential is cuspy.

We plot the final stream stream density using a greyscale just as in
Fig.~\ref{fig:LOG_CHAOTIC_MAP}. 
The axes of the resulting plots in Fig.~\ref{fig:NFW_TRIAXIAL_CHAOS} are the
orbital energy in the triaxial NFW potential and the circularity based on the 
spherical Hernquist profile that was used to generate the initial conditions.
Black points correspond to very large decreases in stream density, hence
to non-regular motion. The upper panel shows all orbits whereas
the lower panel takes only box orbits into account. We distinguished box and tube orbits by
looking at the angular momenta around the symmetry axes. For a box orbit the sign 
of all three momenta changes along the orbit. 
On the other hand a tube orbit has one axis along which the angular momentum 
has a fixed sign.

First of all it is quite striking from these plots that there are orbits that 
are not regular and show up as black points. A line of fixed pericentre 
($1/3 r_s$) in Fig.~\ref{fig:LOG_CHAOTIC_MAP} 
clearly shows that these are orbits that reach the innermost part of the halo,
where they feel the strong triaxiality and the cusp. This is not surprising
because previous studies showed that cuspy, triaxial potentials
exhibit chaos \citep{1998ApJ...506..686V, 2003MNRAS.345..727K}.
These studies found that box orbits are primarily affected but also some tube orbits.
This behaviour can clearly be seen in Fig.~\ref{fig:NFW_TRIAXIAL_CHAOS}

It is not clear how to distinguish between regular and chaotic motion based
on the stream density decrease after a given number of orbital periods. There
is no ``gap" in the stream density distribution which might separate regular and chaotic motion.
Methods like the frequency shift derived from a NAFF analysis  
run into the same problem \citep{1998ApJ...506..686V}.
Lyapunov methods are better suited to this problem,
but may also have problems making a clear distinction \citep{2002CeMDA..82...61K}.
As our objective here is not a precise chaos analysis of our triaxial NFW potential,
we take a fiducial value of $10^{-25}$ for the stream density which separates regular
and chaotic motion. This value is based on the stream density distribution function and corresponds
to a stream density below the ``regular motion bump". Orbits with a stream density
below $10^{-25}$ will be considered as chaotic.
With this criterion about $35 \%$ of the orbits with binding energy
between $0.91~|E_0|$ and $1.00~|E_0|$ are chaotic. This fraction is not particular high
compared to previous studies. \cite{1998ApJ...506..686V}, for example, find fractions up to $80\%$
depending on the parameters they use for their triaxial density profile. Recent studies
of self-consistent models of cuspy triaxial galaxies with dark matter halos \citep{2007ApJ...666..165C}
also find chaotic orbits to play an important role. 
Although we have carried out only a qualitative analysis it
is clear that chaos plays a role in the centre of our model also.
Box orbits are mostly affected by chaotic mixing because they reach the innermost
part of the halo. We note that the four orbits shown in Fig. \ref{fig:TRIAXIAL_FLOWS} 
are regular. None of them has a pericentre distance below the ``critical" distance $1/3 r_s$.

These results show that stream densities near the Sun are predicted to
be much lower for a realistic triaxial potential than for the
corresponding spherical potential. The orbital period near the Sun is
about $2.5\times10^8$~yrs and the distance of the Sun from the
Galactic Centre is roughly 0.2 to $0.3 r_s$.
Figure~\ref{fig:TRIAXIAL_FLOWS} suggests that mixing will have reduced
stream densities from their values at infall by roughly 4 orders of
magnitude for typical streams in the solar neighbourhood.  Since the
mean DM density near the Sun is substantially greater than typical
stream densities at infall, we expect at least $10^5$ CDM streams to
contribute to the density in the solar neighbourhood.  We note that
this is still likely to be a substantial underestimate, as it is based
on a static, smooth halo model and so neglects additional mixing
effects which may be important, in particular mixing in precursor
objects, mixing due to scattering by halo substructure, and chaotic
mixing. Such effects can only be treated properly by applying the GDE
to structure formation in its proper cosmological context.  This
requires the use of N-body simulations. We note that
particles orbiting the innermost part of the halo are also
affected by the disk potential leading to a non-spherical 
contribution to the potential. 

\section{Fine-grained phase-space analysis in N-body codes} \label{sect:N_BODY}

The main motivation of our work is the desire to address issues of
mixing and fine-scale structure in full generality by building the GDE
into current state-of-the-art N-body codes.  We have done this for the
current version of the {\sm GADGET} code \citep{2005MNRAS.364.1105S}.  This
is a massively parallel N-body code which has already been used to
carry out very large cosmological simulations
\citep{2005Natur.435..629S}.  It calculates the gravitational forces
with an efficient TreePM method
\citep{2000ApJS..128..561B,1995ApJS...98..355X, 2002JApA...23..185B}
and uses a domain decomposition scheme based on space-filling
(fractal) Peano-Hilbert curves to achieve good work-load balance in
parallel operation.

To implement the GDE within {\sm GADGET} we needed to extend various parts
of the code. The dynamics of the distortion tensor are driven by the
ordinary gravitational tidal field. The corresponding tidal tensors
for each DM particle have to be calculated using the same Tree- or
TreePM-Method as the forces in order to provide the correct driving
term in the GDE. While the forces are given by the first derivative of
the potential, the tidal tensor is made up of second derivatives. The
particles in an N-body simulation can be thought of as a Monte-Carlo
sampling of the real DM phase-space distribution, but the coarseness
of this sampling introduces unwanted discreteness or ``two-body''
effects which are likely to be more serious for higher derivatives of
the potential.  Such effects are usually mitigated by softening the
gravitational potential of each particle.  {\sm GADGET} uses a spline
softening kernel function with compact support:
\begin{equation}
W_2(u)=
\left\{
\begin{array}{ll}
\frac{16}{3}u^2 - \frac{48}{5} u^4 +\frac{32}{5} u^5 -\frac{14}{5}  ,
&\mbox{$0\le u<\frac{1}{2}$}, \vspace{0.1cm}\\ \nonumber
\frac{1}{15u}+\frac{32}{3}u^2- 16 u^3 +\frac{48}{5}u^4 \\ \nonumber
\mbox{\hspace{1cm}}
-\frac{32}{15}u^5 -\frac{16}{5} ,
& \mbox{$\frac{1}{2}\le u <1$}, \vspace{0.1cm}\\ \nonumber
-\frac{1}{u}, & \mbox{$u \ge 1$}. \\ \nonumber
\end{array}  \nonumber
\right.
\end{equation}
The softened potential of a point mass is then given by
$\Phi^s(\underline{x}) =(G m/h) W_2(|\underline{x}|/h)$ with a
softening length $h=2.8 \epsilon$, where $\epsilon$ is the Plummer
equivalent softening length.  The potential (and so force and tidal
field) become Newtonian if $|\underline{x}| \geq h$. From this
softened potential we can calculate the softened tidal field of a
point mass:
\begin{equation}
T_{ij}^s\left(\underline{x}\right)  = -\frac{\partial^2 \Phi^s\left(\underline{x}\right)}{\partial x_i \partial x_j} = \delta_{ij} m g_1\left(\frac{|\underline{x}|}{h}\right) + x_i x_j m g_2\left(\frac{|\underline{x}|}{h}\right), \nonumber
\label{eq:SOFT_TIDAL} \nonumber
\end{equation}
where
\begin{equation}
g_1(u) = \frac{1}{h^3} 
\left\{ 
\begin{array}{ll}
-\frac{32}{3}+\frac{192}{5}u^2-32u^3, &
\mbox{$u\le\frac{1}{2}$} ,\vspace{0.1cm}\\ \nonumber
\frac{1}{15u^3} - \frac{64}{3} + 48 u \\ \nonumber
\mbox{\hspace{1cm}} - \frac{192}{5}u^2 + \frac{32}{3}u^3, &
\mbox{$\frac{1}{2}<u<1$},\vspace{0.1cm}\\ \nonumber
-\frac{1}{u^3} , & \mbox{$u>1$,}
\end{array} \nonumber
\right.
\label{eq:SOFT_G1}
\end{equation}
and
\begin{equation}
g_2(u) = \frac{1}{h^5} 
\left\{ 
\begin{array}{ll}
\frac{384}{5}- 96 u, &
\mbox{$u\le\frac{1}{2}$},\vspace{0.1cm}\\ \nonumber
-\frac{384}{5} - \frac{1}{5u^5} +\frac{48}{u} +32 u, &
 \mbox{$\frac{1}{2}<u<1$},\vspace{0.1cm}\\ \nonumber 
\frac{3}{u^5}, & \mbox{$u>1$} .
\end{array} \nonumber
\right.
\label{eq:SOFT_G2}
\end{equation}
Thus the softened tidal field acting on particle $k$ at position
$\underline{x}_k$ (its tidal tensor) is given by:
\begin{eqnarray}
T_{ij}\left(\underline{x}_k\right) &=& \sum_{l \neq k}  \Bigg[ \delta_{ij} m_l g_1\left(\frac{|\underline{x}_l-\underline{x}_k|}{h}\right)  \\ \nonumber
&+& \left(x_{l,i}-x_{k,i}\right) \left(x_{l,j}-x_{k,j}\right) m_l g_2\left(\frac{|\underline{x}_l-\underline{x}_k|}{h}\right) \Bigg]\\ \nonumber
&=&\sum_{l} \Bigg[\delta_{ij} m_l g_1\left(\frac{|\underline{x}_l-\underline{x}_k|}{h}\right) \\ \nonumber
&+& \left(x_{l,i}-x_{k,i}\right) \left(x_{l,j}-x_{k,j}\right) m_l g_2\left(\frac{|\underline{x}_l-\underline{x}_k|}{h}\right) \Bigg]\\  \nonumber
&-& \delta_{ij} m_k g_1\left(0\right) 
\label{eq:TOTAL_TIDAL} 
\end{eqnarray}
The last step highlights a difference between the tidal field
calculation and the normal force calculation.  The tidal field is
obtained using the same tree-walk as the forces. The latter are calculated
by evaluating the full sum $\sum_{l}$ without excluding particle $k$.
This is simply to avoid additional bookkeeping; the particle-particle
force vanishes when $l=k$ so including the self-term does not affect
the result. This is not the case for the tidal field, for which one
must add an extra term to the diagonal tidal tensor elements to remove
the self-tidal field. This is similar to the self-energy correction
that is needed when using the tree to evaluate the total potential
energy of the system.

For larger simulations it is not efficient to use the tree alone. In
such cases the TreePM method can be much faster. In this scheme the
potential is split into short-range and long-range parts $\Phi =
\Phi^{\mathrm{short}}+\Phi^{\mathrm{long}}$. Specifically, in Fourier
space {\sm GADGET} takes
\begin{equation} 
\Phi_k^{\mathrm{long}}=\Phi_k \exp\left(-k^2 r_s^2\right)
\label{eq:LONGRANGE}
\end{equation}
where $r_s$ defines the spatial scale of the force split and should
not be confused with the scale radius of the NFW profile.  The
long-range potential is calculated by mesh-based Fourier techniques.
In Fourier space the tidal field can be calculated by just pulling
down $-(i k_j)^2$ with $j=x,y,z$. The short-range potential in real
space is given by:
\begin{equation} 
\Phi_{\mathrm{short}}\left(\underline{x}\right)= \sum_l \Phi^s\left(r_l\right) \mathrm{erfc}\left(\frac{r_l}{2 r_s}\right),
\label{eq:SHORTRANGE}
\end{equation}
and the corresponding short-range part of the tidal field by:
\begin{eqnarray} 
T_{ij}\left(\underline{x}\right)\!\!\!\!  &=& \!\!\!\!\!\!\! \sum_l  \Bigg(T_{ij}^s\left(\underline{x}_l\right) \left(\mathrm{erfc}\left(\frac{r_l}{2 r_s}\right) \nonumber
+ \frac{r_l}{\sqrt{\pi}r_s} \exp\left(-\frac{r_l^2}{4 r_s^2}\right)\right) \\ \nonumber
&-& F_i^s\left(\underline{x}_l\right) \frac{x_{l,j} r_l}{2 \sqrt{\pi} r_s^3} \exp\left(-\frac{r_l^2}{4 r_s^2}\right)  \Bigg), \\ \nonumber
\label{eq:SHORTRANGE_TIDAL} \nonumber
\end{eqnarray}
where $\underline{F}^s$ is the softened point mass force, $x_{l,i}$ is defined as the smallest distance of any of the
periodic images of particle $l$ to the coordinate $x_i$ of $\underline{x}$, and
$r_l=\sqrt{x_{l,x}^2 + x_{l,y}^2 + x_{l,z}^2}$.

The time integration also needs modification in order to integrate the
GDE in parallel with the equations of motion. For this it is
desirable to write both the equations of motion and the GDE in a
time-symmetric way. This fits best into {\sm GADGET}'s
quasi-symplectic integration scheme which is a second-order
leapfrog. For the GDE we need to integrate two differential equations
of second-order to solve for $\underline{\underline{D}}_i$ with
$i=xx,xv$. Let $\underline{\underline{W}}_i$ denote the first time
derivative of $\underline{\underline{D}}_i$. We can then define the
system state vector $\tilde{S}$ as,
\begin{equation}
\tilde{S}=\left(\underline{x},\underline{v},\underline{\underline{D}}_{xx},\underline{\underline{D}}_{xv},\underline{\underline{W}}_{xx},\underline{\underline{W}}_{xv}\right)^\dagger .
\label{eq:SYSTEM_STATE}
\end{equation}
The equations of motion and the GDE can now be written as one equation
for $\tilde{S}$:
\begin{equation}
\ddot{\tilde{S}}\left(t;\overline{x}_0\right) = f\left(\tilde{S}\left(t;\overline{x}_0\right)\right) .
\label{eq:SYSTEM_ODE}
\end{equation}
The right hand side does not depend on the time derivative of
$\tilde{S}$. This allows the use of a time-symmetric leapfrog scheme
with the following Drift- and Kick-operators:
\begin{eqnarray}
D_t(\Delta t) = 
\left\{ 
\begin{array}{llll}
\underline{v} & \rightarrow & \underline{v} & \\
\underline{x} & \rightarrow & \underline{x} + \underline{v} \Delta t & \\
\underline{\underline{W}}_i & \rightarrow & \underline{\underline{W}}_i  &  i=xx,xv \\
\underline{\underline{D}}_i & \rightarrow & \underline{\underline{D}}_i + \underline{\underline{W}}_i \Delta t & i=xx,xv\\
\end{array} 
\right. \\
K_t(\Delta t) =
\left\{ 
\begin{array}{llll}
\underline{x} & \rightarrow & \underline{x} & \\
\underline{v} & \rightarrow & \underline{v} + \underline{F} \Delta t & \\
\underline{\underline{D}}_i & \rightarrow & \underline{\underline{D}}_i  &  i=xx,xv \\
\underline{\underline{W}}_i & \rightarrow & \underline{\underline{W}}_i + \underline{\underline{T}} \Delta t & i=xx,xv\\
\end{array}, 
\right. \\
\label{eq:SYSTEM_KICK} \nonumber
\end{eqnarray}
where $\underline{\underline{T}}$ is the ordinary tidal tensor and
$\underline{F}$ the gravitational force.  Although the time
integration now needs to solve 18 additional non-trivial coupled
second-order differential equations, it turns out that the loss in
performance is not dramatic, even if we do the calculation for all DM
particles in the simulation box.  From a computational point of view,
the strongest impact comes from the extra memory that is needed to
keep track of the distortion tensor. Every particle needs the tidal
tensor (6 numbers due to symmetry) and the distortion tensor (36
numbers).  Nevertheless, with current computer capabilities this is
not a major limitation.

\begin{figure}
\centerline{\includegraphics[width=1.0\linewidth]{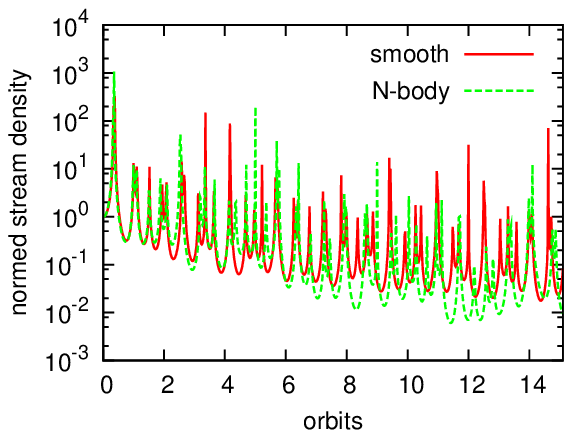}}
\caption{
\label{fig:FLOW_DENSITY}
The stream density evolution found by integrating the GDE along an
orbit in a live N-body realisation of a Hernquist sphere is compared
to that predicted for the same initial condition in the corresponding
analytic potential. The N-body halo was realised with $N=10^5$
particles using a softening length, $\epsilon=1.5$ kpc. The N-body
evolution is very similar in shape and caustic frequency to that in
the smooth potential.  The 6-D phase-space density remained constant
to an accuracy of $10^{-8}$ over the full N-body integration.  }
\end{figure}

In general the state of an N-body simulation is not stored frequently
enough to catch the caustics that occur along each particle's
orbit. To avoid missing these we implemented a caustic finder that
examines every drift operation of the time integration.  As described
above, sign changes in the determinant of the configuration-space
distortion tensor $\underline{\underline{D}}$ indicate that a particle
has passed though a caustic. Whenever this happens the event is
logged. We are then limited only by the time-step of the simulation
and this is normally small enough to catch all large-scale caustics.

For flexibility in testing, we have also implemented the GDE formalism
in a version of {\sm GADGET} which allows certain static potentials,
in particular, NFW, Hernquist and cored ellipsoidal logarithmic
potentials, to be included in addition to the self-gravity of the
particles.

As a first test of our implementation in {\sm GADGET}, we have
compared the behaviour predicted for N-body realisations of a static
Hernquist sphere to that found for an integration in the
corresponding smooth potential. To get a system which resembles the
Milky Way's halo we take $M=1.86 \times 10^{12} \mathrm{M}_{\odot}$
and $a=34.5$ kpc in equation (\ref{eq:HERNQUIST_DENSITY}). The N-body
realisation was constructed as described in section 6.

\begin{figure*}
\centerline{\includegraphics[width=1.0\linewidth]{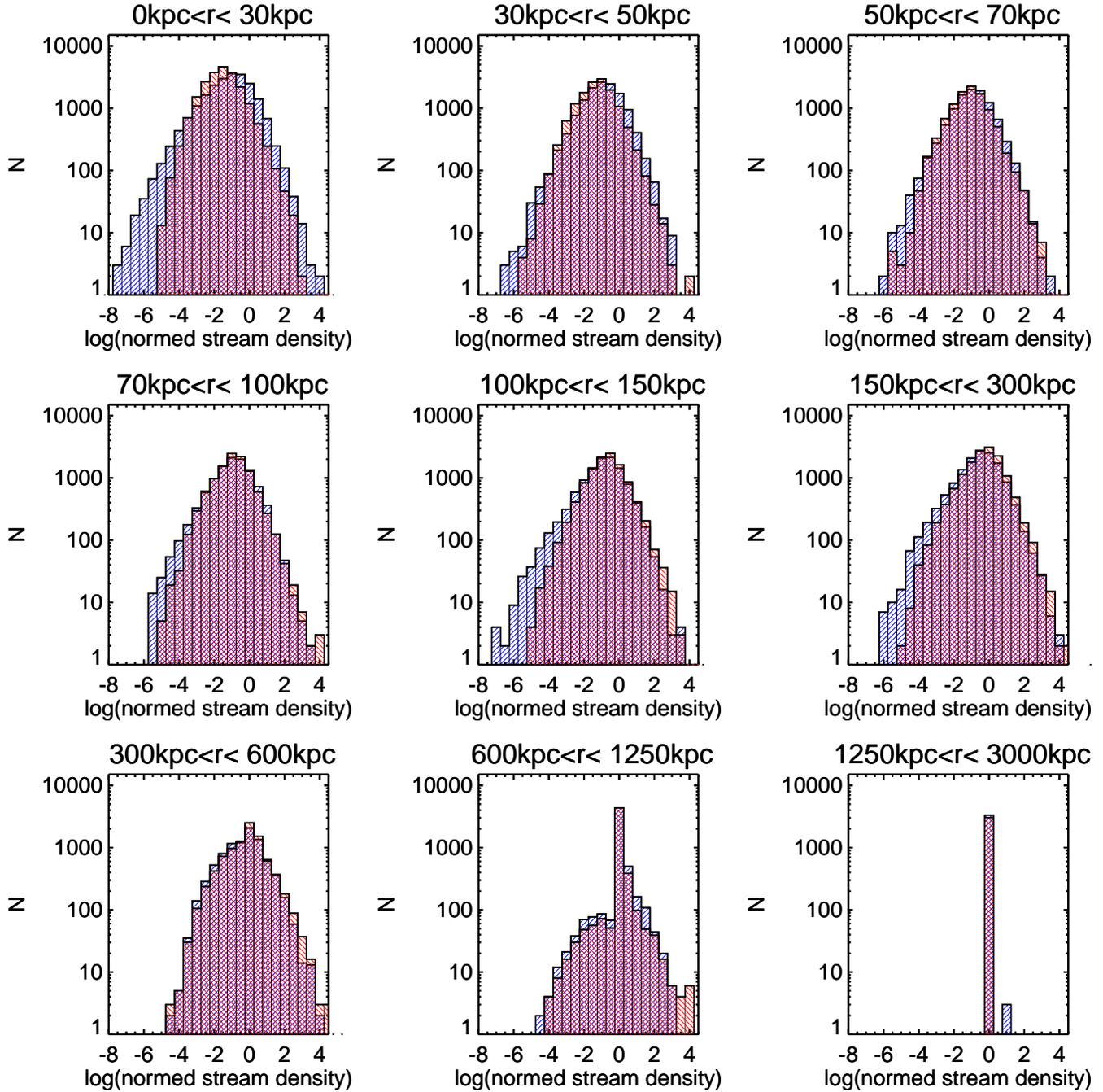}}
\caption{
\label{fig:STREAM_DENSITY_SPHERE}
Stream densities for an $N=10^5$ spherical Hernquist N-body model of the Galactic halo
after $5.0$ Gyr of integration (blue hatching from bottom left to top
right) are compared to those found by integrating from the same
initial conditions in the corresponding smooth potential (red hatching
from bottom right to top left). In each case the sphere was divided
into nine spherical shells and the normed stream densities of all
particles in each shell at the final time were histogrammed with bin
width 0.5 in $\log$(density).  As expected, the lowest stream
densities are reached near the centre of the sphere where the
dynamical time scale is shortest.  In the two outermost spherical shells
most of the particles have not yet undergone many caustics. The stream
density distribution there has a strong bias towards values $>1$.  The
agreement between the N-body live halo and the smooth potential is good.  }
\end{figure*}

In Fig. \ref{fig:FLOW_DENSITY} we compare the evolution of stream
density for a specific orbit in an $N=10^5$ live Hernquist halo to
that found when integrating the same orbit in the
corresponding smooth potential.  The Hernquist sphere had the
parameters given above and the particular orbit chosen here had peri-
and apocentre of $25$ kpc and $33$ kpc, respectively, giving a period of
about 0.5 Gyr. It was integrated for about 15 orbits or 7.5 Gyr.  The
N-body softening was taken to be $\epsilon=1.5$ kpc.  The 6-D
phase-space density remained constant to better than 1 part in $10^8$
in both integrations, but the stream density evolution still differs
significantly between them, in particular in the timing of the
caustics and in the detailed behaviour of the lower envelope. This is
a consequence of the well known divergence between nearby orbits in
N-body systems which is caused by the cumulative effect of many small
perturbations due to discreteness \citep{2003ApJ...585..244K}.  The
GDE is very sensitive to such noise.  The features in the two curves
are, nevertheless very similar, in particular the number and spacing
of caustics and the overall shape.

Fig. \ref{fig:STREAM_DENSITY_SPHERE} shows the normed stream densities
after 5 Gyr of integration for all particles in a live Hernquist halo
with $N=10^5$ and the parameters assumed above.  For this integration
we adopted $\epsilon=2.0$ kpc.  We divide the particles into 9 radial
bins containing approximately equal numbers of particles and then
histogram the stream densities, both for the N-body simulation and for
integrations from the same initial conditions in a smooth
Hernquist potential. Typical stream densities in
Fig. \ref{fig:STREAM_DENSITY_SPHERE} decrease towards the centre of
the sphere. This is because shorter dynamical times result in enhanced
mixing in the inner regions. (Recall that stream densities decrease as
$(t/t_{\mathrm{orbital}})^{-2}$ in a spherical potential, and so are
smallest where the orbital periods are shortest.)  The two outermost
radial shells are dominated by particles with long orbital periods
which, as a result, have stream densities of order unity. The high
stream density tails of the histograms are due to particles which are
close to caustic passage. They thus have the universal power-law shape
$N(>\rho) \propto \rho^{-1}$ expected near a caustic (see, for
example, \cite{2006MNRAS.366.1217M}).

The two sets of histograms in Fig.~\ref{fig:STREAM_DENSITY_SPHERE} are
very similar. Although stream densities evolve differently along
orbits from a given initial condition in the N-body and smooth
potential cases (see Fig.~\ref{fig:FLOW_DENSITY}) the statistical
results for ensembles of initial conditions are similar.  N-body
discreteness effects do not cause substantial {\it systematic} shifts
in the stream density distributions predicted for this test problem. A
small systematic effect is visible at low stream densities. The N-body
integration produces more very low-density streams than the integration
in the corresponding smooth potential. This is indeed due to discreteness effects, as evidenced
by the fact that we find the excess to depend on the N-body softening;
smaller softenings result in a larger tail of extremely low-density
streams. On the other hand, too large a softening leads to incorrect
representation of the mean force near the centre of the system. Thus a
trade-off is needed to define the optimal softening.  This has been
much discussed with reference to conventional N-body simulations
\citep{1996AJ....111.2462M,2000MNRAS.314..475A,2001MNRAS.324..273D,2005ARep...49..470R,2006ApJ...639..617Z}
but we note that the situation is worse for our current application,
since the evolution of our extended state vector
(equation~\ref{eq:SYSTEM_ODE}) depends on the tidal tensor. The
additional spatial derivative relative to the force makes our GDE
integrations substantially more sensitive to discreteness than a
standard N-body integration. This suggests that the optimal choice of
softening will be larger for GDE integrations than for conventional
N-body integrations.

\begin{figure}
\centerline{\includegraphics[width=1.0\linewidth]{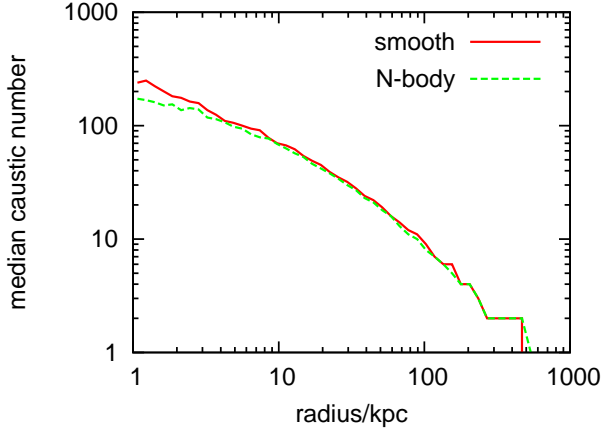}}
\caption{
\label{fig:CAUSTIC_NUMBER}
Number of caustic passages after 5 Gyr as a function of final distance
from the centre for $10^5$ orbits in a Hernquist sphere. The green
dashed curve gives the median number of caustic passages at each
radius for particles in an N-body realisation of the system integrated
with softening parameter, $\epsilon =2$kpc. The red solid curve shows
the result when these same initial conditions are integrated within
the corresponding analytic Hernquist potential. The
results coincide except within about 1.5 softening lengths of the
centre. This demonstrates that caustic counting is very robust against
discreteness effects.  }
\end{figure}

Fig. \ref{fig:STREAM_DENSITY_SPHERE} shows that the high-density tails
of the stream density distribution agree well between the N-body and
smooth potential integrations. This suggests that the number and the strength
of the caustics must be similar in the two cases.  We can check this
explicitly by again dividing the sphere into radial shells and then
calculating the median number of caustic passages by the final time
for the particles which end up in each shell. In
Fig. \ref{fig:CAUSTIC_NUMBER} we compare the results of this exercise
for the N-body and smooth potential integrations using $50$ shells. The level
of agreement is striking. Only within about 3 kpc of the centre is
there a significant difference between the two curves. This is
comparable to the softening used for the N-body system, so it is not
surprising that particles in this inner core pass through fewer
caustics in the N-body case.

\begin{figure}
\centerline{\rotatebox{0}{\includegraphics[width=1\linewidth]{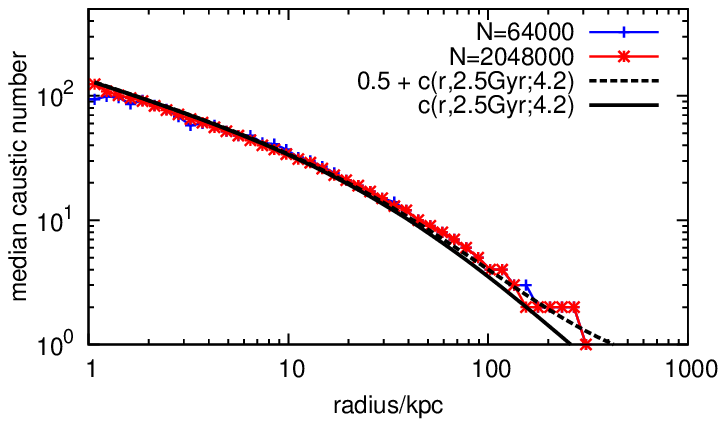}}}
\centerline{\rotatebox{0}{\includegraphics[width=1\linewidth]{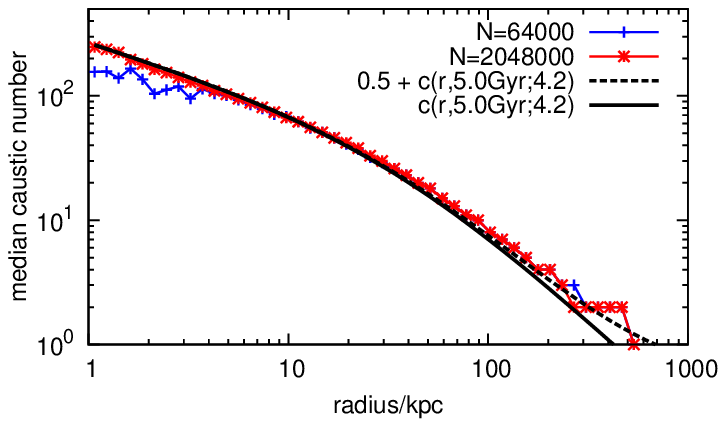}}}
\caption{
\label{fig:CAUSTICS_RADIAL}
Median caustic passage count against radius, as in
Fig.~\ref{fig:CAUSTIC_NUMBER}, but for different particle numbers and
for a smaller softening ($\epsilon = 0.5$~kpc).  The panels show
results after $2.5$ (top) and $5.0$ (bottom) Gyr of evolution.  Clearly
the caustic count goes up between the two times, so the curves are
higher in the lower panel.  The radial dependence is very well
represented by eq. (\ref{eq:CAUSTIC_EST}). Increasing the number of particles by
a factor of $32$ has very little effect on these curves, which also
agree with those of Fig.~\ref{fig:CAUSTIC_NUMBER}. This shows that the
caustic count along an orbit is very insensitive to the N-body
parameters (particle number and softening) used to integrate the
system.}
\end{figure}

The number of caustic passages depends very little N-body parameters.
In Fig.~\ref{fig:CAUSTICS_RADIAL} we plot median caustic count against
radius for two different mass resolutions and for a fixed softening
length of $0.5$ kpc, four times smaller than in
Fig.~\ref{fig:CAUSTIC_NUMBER}. After 5 Gyr, the highest resolution
simulation produces a median caustic count at 1~kpc which agrees with
that for the smooth potential integration in Fig.~\ref{fig:CAUSTIC_NUMBER},
confirming that that the disagreement in that figure was due to the
softening of the N-body simulation. It is remarkable that particle
number has no strong impact on the median caustic count.  The two
simulations in Fig.~\ref{fig:CAUSTICS_RADIAL} differ by a factor of
$32$ in particle mass, yet outside 4~kpc they agree very well both
with each other and with the more softened integration of
Fig.~\ref{fig:CAUSTIC_NUMBER}. The reason for this stability is that the
caustic count is an integer which is augmented only when the
determinant of the distortion tensor changes sign.  As a result, it is
much less sensitive to the exact values of the distortion tensor
elements than is the stream density (which depends on the value of the
determinant).

We can estimate the median number of caustic passages for particles at
radius $r$ very simply as $\kappa t/T(r)$ where $t$ is the age of the
system, $T(r)$ the period of a circular orbit at radius $r$, and
$\kappa$ a proportionality constant.  Then $T(r) = 2 \pi r/V_c(r)$,
where $V_c(r)=\sqrt{GM(r)/r}$ is the circular velocity at radius $r$.
For a Hernquist sphere, the mass $M(r)$ within radius $r$ is $M(r)=M
r^2/(r+a)^2$. Putting all this together we get:
\begin{equation}
 c(r,t; \kappa) = \kappa \frac{t}{T(r)} = \frac{\kappa}{2 \pi} \frac{t \sqrt{G M / r}}{(r+a)}
\label{eq:CAUSTIC_EST}
\end{equation}
where $c(r,t;\kappa)$ is the predicted median caustic count at radius $r$.
This estimate works very well, with a best fit $\kappa=4.2$. The
deviation at large radii where the caustic number is low can be
accommodated simply by adding a small constant offset, as indicated by
the dashed line in the figure. HW (Eq. 37) already showed that
caustics occur in a spherical potential when $p_\theta=0$ or $p_r=0$,
thus at the turning points in the $\theta$ and $r$ coordinates.  If
$p_\theta$ and $p_r$ go through zero at different times we would
expect four caustics per orbital period. This is surprisingly close to
the value of $\kappa$ that we estimate directly from the simulation,
given that the particles seen at radius $r$ actually have a wide range
of orbital periods, rather than all having the circular orbit period
$T(r)$.
  
\begin{figure}
\centerline{\includegraphics[width=1\linewidth]{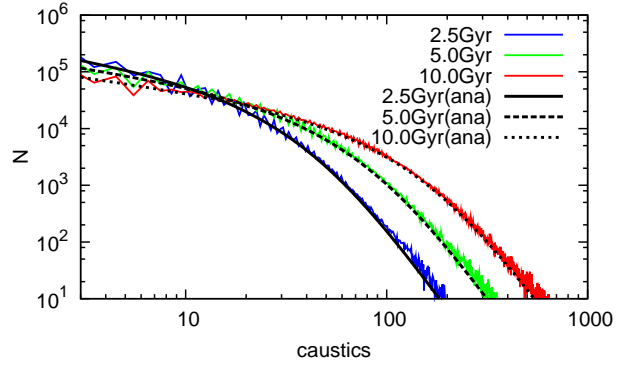}}
\caption{
\label{fig:CAUSTIC_BIN}
The number of particles that have passed through a given number of
caustics is plotted against the number of caustics for our $N=2048000$
simulation after 2.5, 5 and 10~Gyr. The thin black lines are analytic
estimates based on eq.~(\ref{eq:CAUSTIC_BIN_ANA}).  }
\end{figure}

Rather than focussing on the median count of caustic passages as a
function of radius, one can examine the distribution of the number of
caustic passages, i.e. the number of particles that have passed
through a given number of caustics after some given time.
Figure~\ref{fig:CAUSTIC_BIN} shows such distributions after 2.5, 5 and
10 Gyr for our highest resolution ($N=2048000$) simulation. With
increasing time the characteristic number of caustic passages
increases and the number of particles with a small number of caustic
passages decreases.

We can make a simple analytic model for these distributions based on
Eq. (\ref{eq:CAUSTIC_EST}). There are $4 \pi \rho(r) r^2/m\,
\mathrm{d}r$ particles in the interval $(r,r+\mathrm{d}r)$, where $m$
is the mass of a simulation particle and $\rho(r)$ the (analytic)
density profile of the Hernquist sphere.  If we make the approximation
that all particles at radius $r$ have a caustic count equal to the
median count predicted by eq.~(\ref{eq:CAUSTIC_EST}), the number of
particles with caustic counts in the interval $(c(r,t;\kappa),
c(r,t;\kappa) + \mathrm{d}c(r,t;\kappa))$ will be the same as the number
of particles in $(r,r+\mathrm{d}r)$, so:
\begin{eqnarray}
 f(c)\mathrm{d}c &\cong& 4 \pi \rho(r) \frac{1}{m} r^2 \mathrm{d}r \\
 &=&4 \pi \frac{M}{2 \pi} a \frac{r(c,t;\kappa)}{(r(c,t;\kappa)+a)^3} \frac{1}{m}
\label{eq:CAUSTIC_BIN_ANA}
 \left|\frac{\mathrm{d}r(c,t;\kappa)}{\mathrm{d}c} \right|\mathrm{d}c, 
\end{eqnarray}
where $r(c,t;\kappa)$ is the inverse function of $c(r,t;\kappa)$. As
Fig.~\ref{fig:CAUSTIC_BIN} shows, this formula represents the
simulation results very well, suggesting that the variation in caustic
count with radius is more important than the scatter in caustic count
at given radius for determining the overall shape of the count
distribution.

From these first tests we conclude that our N-body implementation is
working well, that caustic properties can be predicted very robustly,
at least when the caustics reflect large-scale structure in the
system, and that stream densities can also be predicted reliably
provided care is taken to ensure that discreteness effects are under
control.

\section{Conclusion} \label{sect:discussion}

Direct DM detection experiments operate on length-scales far below the
resolution of current structure-formation simulations.  The
fine-grained phase-space structure on these scales will determine the
signal they see. In addition, small-scale structure can substantially
enhance the annihilation signal that is the target of current indirect
detection experiments. A better understanding of such structure within
the current concordance $\Lambda$CDM cosmology is thus critical for
analysing and interpreting all current DM searches.

We propose a new route to tackle these issues. Rather than improving
simulations simply by increasing the number of particles, we attach
additional information to each particle, namely a phase-space
distortion tensor which allows us to follow the evolution of the
fine-grained phase-space distribution in the immediate neighbourhood
of the particle.  We introduce the Geodesic Deviation Equation (GDE)
as a general tool for calculating the evolution of this distortion
along any particle trajectory.  The projection from phase-space to
configuration-space yields the density of the particular CDM stream
that particle is embedded in and can also identify when the particle
passes through a caustic.

This technique makes the fine-grained phase-space structure
accessible. It enables studies of the phase-space structure of general
non-integrable static potentials which reproduce all the results
previously obtained using frequency analysis methods, identifying
chaotic regions and finding substructure in regular regions in the
form of resonances.  In addition, it can be used to quantify mixing
rates and to locate caustics.  We demonstrate these capabilities for
the complex phase-space structure of the ellipsoidal logarithmic
potential with a core. All relevant phase-space regions could be
identified by solving the GDE along the orbit. We have written a code,
DaMaFlow, that allows us to carry out such stream density analyses for
a wide variety of potentials in a very efficient way.

Stream density evolution is very sensitive to the shape of the
underlying potential. We demonstrate this by comparing results for a
realistic CDM halo with radially varying shape to those for a
spherical halo with similar radial density profile. After 100 orbits
the predicted stream densities in the inner regions differ by a factor
of 100.  In general we expect the stream densities to decrease as
$(t/t_{\mathrm{orbital}})^{-3}$ for regular orbits and even faster for
chaotic orbits, rather than as $(t/t_{\mathrm{orbital}})^{-2}$, the
result found for orbits in a spherical potential.  Scaling to the
Milky Way leads us to estimate that there should be at least $10^5$
streams passing through the solar system.

The potentially revolutionary advantage of our approach, and our main
reason for pursuing it, is that it applies equally well to
non-symmetric, non-static situations of the kind that generically
arise in CDM cosmologies. Indeed, it can be implemented in a
relatively straightforward way in current state-of-the-art
cosmological N-body codes.  We have carried out such an implementation
in the {\sm GADGET} code and have presented some tests based on
equilibrium Hernquist models.  The N-body implementation is able to
conserve 6-D phase-space density to high accuracy along individual
particle orbits. In addition, it qualitatively reproduces the results
found in the corresponding smooth potential for the evolution of stream density along individual
orbits, and it reproduces the statistical results found for ensembles
of orbits to impressive accuracy. The identification of caustic
passages is particularly robust, showing very little dependence on
N-body parameters such as particle number and softening. Thus
discreteness effects appear to be well under control, at least for the
large N systems studied here. The remarkably robust identification of
caustic properties makes us optimistic that we will be able to
calculate annihilation boost factors due to caustics in fully
realistic $\Lambda$CDM dark halos.

In future applications we will use these techniques to address mixing
and DM detection issues within fully general simulations of the
$\Lambda$CDM structure formation model.  

\section*{acknowledgements}
M.V. thanks Ernst Hairer for helpful discussions on geometrical integrators.
\bibliography{lit}

\label{lastpage}
\end{document}